\documentclass[aps,pre,twocolumn,citeautoscript,superscriptaddress,byrevtex,nofootinbib,nobalancelastpage,floatfix]{revtex4}
\usepackage[pdftex]{graphicx}         
\usepackage{fancyhdr}                 
\usepackage{amsmath}                  
\usepackage{bm}                       
\usepackage{textcomp}
\usepackage[colorlinks=true,
    citecolor=blue,
    linkcolor=blue,
    urlcolor=blue,
    pagebackref=false]{hyperref}
\usepackage{bm}         
\usepackage{setspace}
\usepackage[thickspace,amssymb,detect-all]{SIunits}
\usepackage{multirow}
\usepackage[compact]{titlesec}
\usepackage{palatino}  
\usepackage{cases}

\begin{document}

    \pagestyle{fancy}

        \lhead{\footnotesize \textsf{}}
        \chead{\normalsize }
        \rhead{\footnotesize \textsf{\thepage}}
        \lfoot{}
        \cfoot{}
        \rfoot{}

    \title{Dynamic nuclear polarization in a magnetic resonance force microscope experiment}

    \author{Corinne E. Isaac}
    \author{Christine M. Gleave}
    \author{Pam\'{e}la T. Nasr}
    \author{Hoang L. Nguyen}
    \author{Elizabeth A. Curley}
    \affiliation{Department of Chemistry and Chemical Biology,
    	Ithaca, New York 14853-1301, USA}
    \author{Jonilyn L. Yoder}
    \altaffiliation{n\'{e}e Longenecker.
    Present address: Quantum Information and Integrated Nanosystems Group,
    Lincoln Laboratory, Massachusetts Institute of Technology,
    Lexington, Massachusetts 02420-9108, USA}
    \affiliation{Department of Chemistry and Chemical Biology,
    	Ithaca, New York 14853-1301, USA}
    \author{Eric W. Moore}
    \altaffiliation{Present address: Department of Chemistry,
    	Lehigh University, Bethlehem, Pennsylvania 18015, USA}
    \affiliation{Department of Chemistry and Chemical Biology,
    	Ithaca, New York 14853-1301, USA}
    \author{Lei Chen}
    \altaffiliation{Present address: State Key Laboratory of Functional
    	Materials for Informatics, Shanghai Institute of Microsystem
		and Information Technology, Chinese Academy of Sciences, Shanghai, China}
    \affiliation{Department of Chemistry and Chemical Biology,
    	Ithaca, New York 14853-1301, USA}
    \author{John A. Marohn}
    \affiliation{Department of Chemistry and Chemical Biology,
    	Ithaca, New York 14853-1301, USA}

\begin{abstract}
We report achieving enhanced nuclear magnetization in a magnetic resonance force microscope experiment at 0.6~tesla and 4.2~kelvin using the dynamic nuclear polarization (DNP) effect.
In our experiments a microwire coplanar waveguide delivered radiowaves to excite nuclear spins and microwaves to excite electron spins in a 250~nm thick nitroxide-doped polystyrene sample.
Both electron and proton spin resonance were observed as a change in the mechanical resonance frequency of a nearby cantilever having a micron-scale nickel tip.
NMR signal, not observable from Curie-law magnetization at 0.6~\tesla, became observable when microwave irradiation was applied to saturate the electron spins.
The resulting NMR signal's size, buildup time, dependence on microwave power, and dependence on irradiation frequency was consistent with a transfer of magnetization from electron spins to nuclear spins.
Due to the presence of an inhomogenous magnetic field introduced by the cantilever's magnetic tip, the electron spins in the sample were saturated in a microwave-resonant slice 10's of nm thick.
The spatial distribution of the nuclear polarization enhancement factor $\epsilon$ was mapped by varying the frequency of the applied radiowaves.
The observed enhancement factor was zero for spins in the center of the resonant slice, was $\epsilon = +10$ to $+20$ for spins proximal to the magnet, and was $\epsilon = -10$ to $-20$ for spins distal to the magnet.
We show that this bipolar nuclear magnetization profile is consistent with cross-effect DNP in a $\sim \! 10^{5} \: \tesla \: \meter^{-1}$ magnetic field gradient.
Potential challenges associated with generating and using DNP-enhanced nuclear magnetization in a nanometer-resolution magnetic resonance imaging experiment are elucidated and discussed.
\end{abstract}

\date{\today}
\maketitle

We report achieving enhanced nuclear magnetization in a magnetic resonance force microscope experiment at 0.6~tesla and 4.2~kelvin using the dynamic nuclear polarization (DNP) effect.
In our experiments a microwire coplanar waveguide delivered radiowaves to excite nuclear spins and microwaves to excite electron spins in a 250~nm thick nitroxide-doped polystyrene sample.
Both electron and proton spin resonance were observed as a change in the mechanical resonance frequency of a nearby cantilever having a micron-scale nickel tip.
NMR signal, not observable from Curie-law magnetization at 0.6~\tesla, became observable when microwave irradiation was applied to saturate the electron spins.
The resulting NMR signal's size, buildup time, dependence on microwave power, and dependence on irradiation frequency was consistent with a transfer of magnetization from electron spins to nuclear spins.
Due to the presence of an inhomogenous magnetic field introduced by the cantilever's magnetic tip, the electron spins in the sample were saturated in a microwave-resonant slice 10's of nm thick.
The spatial distribution of the nuclear polarization enhancement factor $\epsilon$ was mapped by varying the frequency of the applied radiowaves.
The observed enhancement factor was zero for spins in the center of the resonant slice, was $\epsilon = +10$ to $+20$ for spins proximal to the magnet, and was $\epsilon = -10$ to $-20$ for spins distal to the magnet.
We show that this bipolar nuclear magnetization profile is consistent with cross-effect DNP in a $\sim \! 10^{5} \: \tesla \: \meter^{-1}$ magnetic field gradient.
Potential challenges associated with generating and using DNP-enhanced nuclear magnetization in a nanometer-resolution magnetic resonance imaging experiment are elucidated and discussed.

\section{Introduction}

Magnetic resonance force microscopy (MRFM) is a highly sensitive method for detecting and imaging magnetic resonance \cite{Kuehn2008feb,Poggio2010aug}.
The highest-resolution MRFM imaging experiment to date achieved a spatial resolution of 4 to 10~nm \cite{Degen2009feb}, on the verge of what is necessary to study individual macromolecular complexes, but this experiment required that the sample be affixed to a high compliance microcantilever.
Proton magnetic resonance was observed in a polymer film at comparable sensitivity in a scanned-probe experiment employing a magnet-tipped cantilever \cite{Longenecker2012nov}, suggesting the possibility of performing a nanometer-resolution magnetic resonance imaging (nano-MRI) experiment on an as-fabricated device or a flash-frozen biological sample.
Remarkably, these experiments detected magnetic resonance as a modulation of statistical fluctuations in the sample's proton magnetization.
In a small spin ensemble, these random-sign, statistical ``spin noise'' fluctuations in magnetization greatly exceed the thermal equilibrium, Curie-law magnetization that one usually observes in a magnetic resonance experiment.
Here we use dynamic nuclear polarization (DNP) to create hyperthermal nuclear spin magnetization in an MRFM experiment, with the goal of pushing the experiment out of the spin-noise limit.

The ultimate goal of this work is to increase the signal-to-noise ratio (SNR) of the MRFM experiment.  The SNR enhancement achievable with DNP in an inductively-detected magnetic resonance experiment is determined primarily by the ratio of the hyperpolarized magnetization to the thermally-polarized magnetization.  Assessing the SNR achievable with DNP in an MRFM experiment is not so simple.  Because of the small spin ensembles observed in an MRFM experiment, at low polarization it is preferable to detect magnetization fluctuations, while at high polarization detecting the average magnetization gives higher SNR.  To assess the DNP gain in MRFM, one therefore needs to compare the SNR of two very different experiments.

\begin{figure}
	\centering
	\includegraphics[width=3.50in]{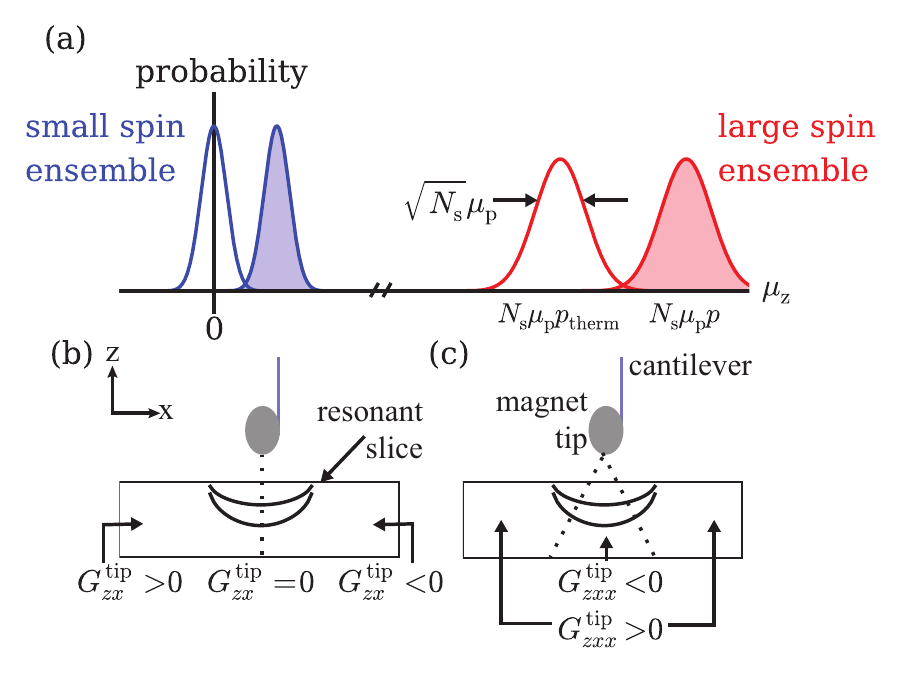}
    \caption{Detecting DNP-enhanced nuclear magnetization in a magnetic resonance force microscope experiment. (a) Magnetization distribution for $N_{\text{s}}$ protons in a small ensemble (blue curve) and large ensemble (red curve) in an external magnetic field, without DNP (unshaded curve) and with DNP (shaded curve).  (b) Detecting magnetic resonance as a  force acting on the cantilever: cantilever, resonant slice of magnetization, and tip-field gradient.  Because of the symmetry of the gradient, the net force on the cantilever from uniformly polarized spins is \emph{zero}. (c) Detecting magnetic resonance as a cantilever frequency shift: cantilever, resonant slice of magnetization, and the tip-field second derivative. \label{Fig:spin_polarization_and_detection}}
\end{figure}

Before continuing, it is helpful to consider the ratio of magnetization fluctuations to the thermal-average magnetization.  The thermal Curie-law magnetization is
\begin{equation}
	\mu_z
		= N_{\text{s}} \, \mu_{\text{p}} \, p_{\text{therm}},
	\label{Eq:muz-mean}
\end{equation}
where $N_{\text{s}}$ is the number of spins in the sample, assumed here to be protons; $\mu_{\text{p}} = 1.41 \times 10^{-26} \: \joule \: \tesla^{-1}$ is the proton magnetic moment; and $p_{\text{therm}} = \tanh{(\mu_{\text{p}} B_0 / k_{\text{B}} T_0)}$ is the thermal spin polarization, with $B_0$ the external magnetic field, $T_0$ the temperature, and $k_{\text{B}}$ Boltzmann's constant.
There is a statistical uncertainty in the spin magnetization whose root-mean-square variation is given by
\begin{equation}
	\delta \mu_z^{\text{rms}}
		=  \sqrt{N_{\text{s}}} \, \mu_{\text{p}} \: \sqrt{1 - p^2_{\text{therm}}}.
	\label{Eq:muz-std}
\end{equation}
At thermal equilibrium, the probability of measuring a certain magnetization is described by a Gaussian distribution whose mean is given by Eq.~\ref{Eq:muz-mean} and whose standard deviation is given by Eq.~\ref{Eq:muz-std} (Fig.~\ref{Fig:spin_polarization_and_detection}(a)).
In a large ensemble the Curie-law magnetization exceeds the root-mean-square variation, while in a small ensemble the root-mean-square variation exceeds the Curie-law magnetization.
In most experiments $p_{\text{therm}} \ll 1$, and in this limit the crossover from large-ensemble to small-ensemble behavior occurs when $N_{\text{s}} \leq 1/p_{\text{therm}}^2$.

Dynamic nuclear polarization increases the spin polarization to
\begin{equation}
p = \epsilon \: p_{\text{therm}}
\end{equation}
and increases the magnetization to $\mu_z = p \, N_{\text{s}} \, \mu_{\text{p}}$, with $\epsilon$ an enhancement factor that lies between $-660$ and $+660$ for protons.
The time-averaged net magnetization now exceeds the root-mean-square magnetization fluctuations when
\begin{equation}
	N_{\text{s}} \geq \frac{1}{p^2}
		\Leftrightarrow
		p \sqrt{N_{\text{s}}} \geq 1.
	\label{Eq:p-crossover}
\end{equation}
By increasing $p$, DNP allows us to study smaller ensembles of spins while remaining in a regime where the average magnetization dominates over the magnetization fluctuations.

Now let us assess the potential signal-to-noise ratio achievable in a magnetic resonance force microscope experiment enhanced by DNP.
Degen and coworkers carefully considered the SNR for detecting magnetization fluctuations from $N_{\text{s}}$ protons in an MRFM experiment \cite{Degen2007dec}.
Fluctuating proton magnetization gives rise to a stochastic force acting on the cantilever whose variance is
\begin{equation}
	\sigma_{\text{spin}}^{2}
	= N_{\text{s}} \mu_{\text{p}}^2 (G^{\text{tip}}_{zx})^2
	= N_{\text{s}} F_1^{2}
	\label{Eq:sigma2spin}
\end{equation}
where $G^{\text{tip}}_{zx} = \partial B^{\text{tip}}_{z}/\partial x$ is the field gradient from the magnetic tip at the location of the spins.
In the above equation we have written the variance in terms of the magnitude of force from a single proton, $F_1 = \lvert \mu_{\text{p}} G^{\text{tip}}_{zx}\rvert$.  The challenge is to perform enough measurements to determine this variance with certainty.
In the MRFM experiment, radiofrequency (rf) waves are applied to cyclicly invert the proton spins twice per cantilever period, creating a cantilever-resonant spin force.
In the presence of the rf, the magnetization fluctuations exhibit a correlation time $\tau_{\text{m}}$ that is typically shorter than $T_1$, the spin-lattice relaxation time, but longer than $T_{1\rho}$, the spin-lattice relaxation time in the rotating frame.
Degen and coworkers showed that the magnetization fluctuations could be detected with improved SNR by applying rf pulses every $t$ seconds to actively randomize the sample magnetization.
Ideally, $t < \tau_{\text{m}}$, allowing the acquisition of more independent measurements of the sample's magnetization than without the periodic randomization.
Decrease the measurement time too much, however, and cantilever mean-square force fluctuations begin to obscure the spin signal.  There is an optimal reset time given by \cite{Degen2007dec}
\begin{equation}
	t_{\text{r}}^{\text{opt}}
	= \frac{P_{\delta F}}
	       {\sqrt{2} \, \sigma_{\text{spin}}^{2}}
	\label{Eq:tropt}
\end{equation}
with
\begin{equation}
	P_{\delta F} = 4 k_{\text{B}} T_0 \, \Gamma
	\label{Eq:PdF}
\end{equation}
the power spectrum of environmental force fluctuations acting on the cantilever and $ \Gamma = k_{\text{c}}/(2 \pi \, f_\text{c} \, Q)$ the cantilever dissipation constant at a temperature $T_0$.
The SNR for detecting magnetization fluctuations may be written as\cite{Degen2007dec}
\begin{equation}
\text{SNR}_{1}
	=
	\alpha_1 \! \! \; ( r )
	\times
	\sqrt{ \frac{T_{\text{acq}}}{t_{\text{r}}^{\text{opt}}} }
	\label{Eq:SNR1}
	\end{equation}
with $T_{\text{acq}}$ the total signal-acquisition time,
\begin{equation}
	r = \frac{t\hfill}
	         {t_{\text{r}}^{\text{opt}}}
	\text{, and }
	\alpha_{1}(r) =
		\sqrt{\frac{r}{2 + 2 \sqrt{2} \, r + 2 \, r^2}}.
	\label{Eq:alpha1}
\end{equation}
Here $t$ is the measurement time and $\alpha_{1}$ is a unitless constant that depends weakly on the ratio $r$ of the measurement time to the optimal reset time.
With $t$ adjusted to be optimal, $r = 1$ and $\alpha_1 = 0.38$.

Now let us consider the SNR for detecting the average magnetization in a magnetic resonance force microscope (MRFM) experiment.
For simplicity, let each experiment begin with the average magnetization equal to zero and let the sample polarize for a time $T$ before detecting.
The SNR for this polarized-spin experiment is derived in the Appendix. The result is
\begin{equation}
	\text{SNR}_{2}
	= \frac
	  {p N_{\text{s}} F_1
	  	(1 - e^{-T/T_1})
		\dfrac{\tau_{\text{m}}}{2 t}
		(1 - e^{-2 t / \tau_{\text{m}}})
	  }
	  {\sqrt{
	  	\dfrac{T + t}{T_{\text{acq}}}
		\left(
			\dfrac{P_{\delta F}}{4 t} + \sigma_{\text{spin}}^2
		\right)}
	  }.
	\label{Eq:SNR2-symm}
\end{equation}
The measurement is {\it detector-noise limited} when $P_{\delta F} \gg 4 \, t \, \sigma_{\text{spin}}^2$, or equivalently $t \ll t_{\text{r}}^{\text{opt}}$.
In this limit, for $t \ll T$, SNR$_2$ is maximized by setting $T \approx \, 1.256 T_1$ and $ t \approx 0.628
\, \tau_{\text{m}}$.
In the extreme case that $\tau_\text{m} \rightarrow T_1$, SNR$_2$ is maximized by setting $T \approx \, 1.655 T_1$ and $ t \approx 0.461  \, T_1$.
In the detector-noise limit, the signal-to-noise ratio for the polarized-spin experiment may be summarized as
\begin{subnumcases}
{\text{SNR}_2
	\approx
	\dfrac{p N_{\text{s}} F_1}
	            {\sqrt{P_{\delta F}/T_{\text{acq}}}}
	\times
	\label{Eq:SNR2}
}
	  0.58 \left( \dfrac{\tau_{\text{m}}}{T_1} \right)^{1/2}
	       & \hspace{-0.25in} $\tau_{\text{m}} \ll T_1$ \label{Eq:SNR2-symm:fast} \\
	  0.49 & \hspace{-0.25in} $\tau_{\text{m}} \rightarrow T_1$ \label{Eq:SNR2-symm:slow}
\end{subnumcases}

Our motivation for pursuing DNP in an MRFM experiment is revealed by comparing the SNR in the polarized-spin experiment, Eq.~\ref{Eq:SNR2-symm}, to the SNR in the unpolarized-spin experiment, Eq.~\ref{Eq:SNR1}.
An analytical result for this ratio can be obtained in two limiting cases.
Above we considered Eq.~\ref{Eq:SNR2-symm} in the detector-noise limit.
The measurement is {\it spin-noise limited} when $P_{\delta F} \ll 4 \, t \, \sigma_{\text{spin}}^2$, or equivalently $t \gg t_{\text{r}}^{\text{opt}}$.
For $t \ll T$, SNR$_2$ is now maximized by setting $T \approx 1.256 \, T_1$, and keeping $t \ll \tau_{\text{m}}$.

The signal to noise ratio in these two limits is
\begin{equation}
    \frac{\text{SNR}_2}{\text{SNR}_1}
    = p \sqrt{N_{\text{s}}}
    \times \begin{cases} \dfrac{0.48}{\alpha_1}\sqrt{\dfrac{\tau_{\text{m}}}{T_1}} & {\text{\it detector-noise limit}} \\
    \dfrac{0.64}{\alpha_1}\sqrt{\dfrac{t_{\text{r}}^{\text{opt}}}{T_1}} & {\text{\it spin-noise limit}}
    \end{cases}
\end{equation}
We see that in both cases the SNR of the polarized-spin experiment is larger than that of the unpolarized-spin experiment by $p \sqrt{N_{\text{s}}}$ times a numerical factor.
The potentially significant numerical factor depends on the relaxation times of the sample and the measurement sensitivity expressed in terms of Degen's optimal reset time $t_{\text{r}}^{\text{opt}}$.
In the limit of slow modulation $\tau_{\text{m}}$ approaches $T_1$ and $\text{SNR}_{2\text{A}}/\text{SNR}_{1} \geq 1$ when $p \sqrt{N_{\text{s}}} \geq 1$ --- precisely the spin-noise-avoidance criterion introduced in Eq.~\ref{Eq:p-crossover}.

In addition to improving SNR, there are a number of other reasons for wanting to detect signal from well-polarized sample spins in an MRFM experiment.
Imparting the nanoscale ensemble of spins detected in an MRFM experiment with a non-zero net spin polarization will facilitate the detection of dilute species via polarization transfer \cite{Eberhardt2007may} and double resonance \cite{Lurie1964feb}.
Moreover, the ability to create and detect a net spin polarization in a nanometer-scale sample is expected to increase the resolution of imaging experiments.
In the virus imaging experiment of Ref.~\citenum{Degen2009feb}, Degen and co-workers collected a force-noise map while slowly scanning the sample with respect to the magnet.
They subsequently applied a time-consuming, iterative non-linear deconvolution to the force-noise map to reconstruct an image of the sample's spin density.
Nichol and Budakian showed that a spin-density map could instead be obtained by evolving spin fluctuations in a pulsed magnetic field gradient; they built up a multi-dimensional correlation function through signal averaging and applied a Fourier transform (FT) to obtain an image \cite{Nichol2013sep}.
While this FT approach has many advantages including rapid and essentially linear image reconstruction and a favorable signal-to-noise ratio, obtaining an FT image in the spin-noise limit requires an inordinate amount of signal averaging.
The experiments described below were motivated by our conclusion that the time required to perform an FT-MRFM imaging experiment on a nanoscale ensemble of spins could be decreased significantly by using DNP to create a magnetic resonance signal with a well-defined sign.

In an inductively-detected magnetic resonance experiment, the Curie-law magnetization and the DNP-enhanced magnetization are both detected as a Faraday-law voltage.
In an MRFM experiment, detecting DNP-enhanced nuclear magnetization requires some thought.
The most sensitive MRFM experiments to date \cite{Rugar2004jul,Longenecker2012nov,Degen2009feb,Nichol2012feb} have detected resonance-induced modulations of longitudinal spin magnetization as a change in the force acting on a cantilever,
\begin{equation}
	{\Delta F}_{\text{spin}}(t)
		= \sum_{k} \Delta \mu_{z,k}(t) \, G_{zx}^{\text{tip}} (\bm{r}_k,\,t).
		\label{Eq:force_detection}
\end{equation}
Here $z$ is the direction of the applied magnetic field, $G_{zx}^{\text{tip}} = \partial B^{\text{tip}_z}/\partial x$ is the derivative of the tip magnetic field in the direction $x$ of the cantilever motion, $\bm{r}_k$ is the location of the $k^{\text{th}}$ spin, and the sum is over all spins in resonance.
To avoid snap-in to contact, the cantilever is operated in the ``hang-down'' geometry \cite{Mamin2003nov}.
Due to the symmetry of the gradient, Fig.~\ref{Fig:spin_polarization_and_detection}(b), the net force from uniformly polarized spins is \emph{zero} in this geometry.\footnote{Spin fluctuations create a temporary left/right imbalance in magnetization observable as a force fluctuation in the experiment of Fig.~\ref{Fig:spin_polarization_and_detection}(b)~\cite{Degen2007dec}.}
The gradient in the spin force shifts the resonance frequency of the cantilever by an amount
\begin{equation}
	{\Delta f}_{\text{spin}}(t)
		= \frac{f_c}{2 k_c} \sum_{k} \Delta \mu_{z,k}(t) \, G_{zxx}^{\text{tip}} (\bm{r}_k),
		\label{Eq:freq_shift}
\end{equation}
where $f_c$ is the cantilever frequency, $k_c$ is the cantilever spring constant, $G_{zxx}^{\text{tip}} = \partial^2 B^{\text{tip}_z}/\partial x^2$, and the sum is over all spins in resonance.
As can be seen in Fig.~\ref{Fig:spin_polarization_and_detection}(c), the second derivative of the tip field $G_{zxx}^{\text{tip}}$ is a symmetric function of $x$.
Consequently, the frequency shift given by Eq.~\ref{Eq:freq_shift} is sensitive to the average, net magnetization \cite{Garner2004jun,Moore2009dec,Alexson2012jul} as well as magnetization fluctuations \cite{Mamin2007may}.
In the experiments detailed below, we detect changes in sample magnetization due to magnetic resonance and DNP as a shift in the resonance frequency of a magnet-tipped cantilever.

\begin{figure}
	\centering
    \includegraphics[width=3.50in]{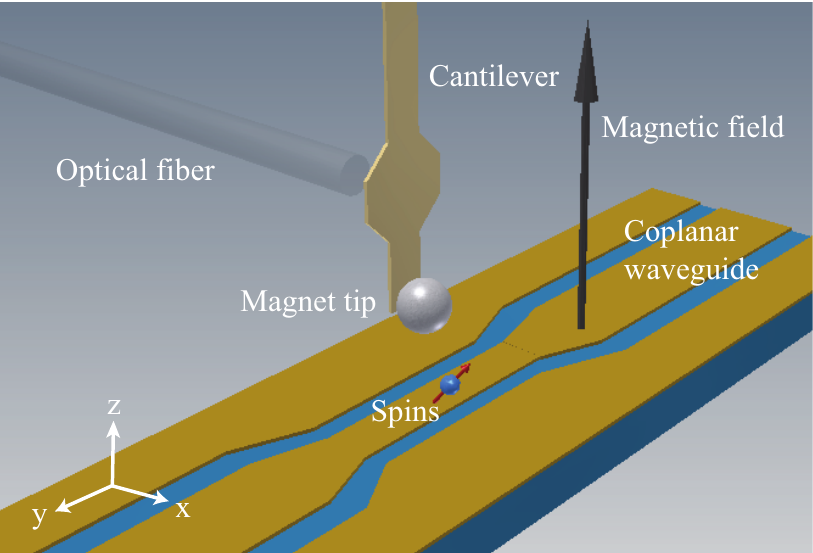}
    \caption{Schematic of the experiment. A $7 \, \micro \meter$ diameter nickel tip was affixed to the end of a silicon cantilever; the magnet-tipped cantilever was brought over the center conductor of a $10 \micro\meter$ wide coplanar waveguide; the waveguide was coated with a $250 \, \nano\meter$ thick film of 40~mM TEMPAMINE in polystyrene. A polarizing magnetic field was applied in the direction of the cantilever's long axis ($z$, black arrow) and the cantilever oscillated in the $x$ direction.  An optical fiber was used to monitor the cantilever motion.  For clarity, the substrate and cantilever are not drawn to scale. }
    \label{Fig:experimental_setup}
\end{figure}

There are only a few examples of observing hyperthermal spin polarization in an MRFM experiment.
Thurber, Smith, and coworkers used optical-pumping DNP to increase the nuclear spin magnetization 12-fold in a gallium arsenide sample affixed to a cantilever in a magnetic resonance force microscope experiment \cite{Thurber2003jun, Thurber2002mar}.
Optical pumping is challenging to implement in an MRFM experiment because of heating, and the optical nuclear polarization mechanism is restricted to semiconducting samples.
Chen, Marohn, and coworkers observed a long-lived shift in the frequency of a magnet-tipped cantilever in an ESR-MRFM experiment carried out on a nitroxide-doped perdeuterated polystyrene film \cite{Chen2013apr,Chen2013jul}.
The size and buildup time of the frequency shift signal led them to hypothesize that it arose from a DNP enhancement of $^{2}\text{H}$ magnetization; they were unable to definitively prove this hypothesis, however, because they were not able to apply the radiowaves required to flip the $^{2}\text{H}$ nuclear spins.

Below we use DNP to create enhanced $^{1}\text{H}$ magnetization in a magnet-on-cantilever MRFM experiment carried out on a nitroxide-doped polystyrene film at 4.2 kelvin.
The experiment is sketched in Fig.~\ref{Fig:experimental_setup}.
In these experiments, we can simultaneously apply both microwaves and radiowaves to the sample and thereby verify that DNP-enhanced magnetization has been created, can measure the background Curie-law signal and therefore quantify the enhancement, and can vary the rf center frequency to probe the spatial distribution of the enhanced nuclear magnetization.
There are only a few examples of observing DNP at liquid helium temperatures or in the TEMPAMINE/polystyrene system studied here  \cite{Vandenbrandt2004jun,Cho2007aug,Siaw2012aug}.  We know of no precedent for observing DNP in the large magnetic field gradient present in our MRFM experiment.

\section{Materials and Methods}
\label{Sec:MM}

\subsection{Sample.}

The sample was a 250 nm thick film of 40 mM TEMPAMINE (4-amino-2,2,6,6-tetramethylpiperidine-1-oxyl; Sigma, 14691-88-4) in polystyrene (Scientific Polymer, 282639, $M_n$ = 139.5 $\times$ 10$^3$ and $M_w/M_n$ = 1.09).  The film was prepared and spin cast onto the coplanar waveguide described below following the protocols given in the Supporting Information of Ref.~\citenum{Moore2009dec}.

\subsection{Cantilever.}

Magnetic resonance was detected with a custom-fabricated attonewton-sensitivity silicon cantilever, as described in Refs.~\citenum{Jenkins2004may} and~\citenum{Hickman2010nov}.
A radius $r = 3.5 \micro \meter$ nickel sphere (saturation magnetization $\mu_{\text{0}} \, M$ = 0.6 T) was manually affixed to the leading edge of the cantilever with epoxy.

Cantilever displacement was observed with a temperature-tuned fiber optic interferometer \cite{Bruland1999sep} (wavelength $\lambda = 1310 \: \nano\meter$).
The power spectrum of cantilever displacement fluctuations was recorded and a spring constant $k_{\text{c}}$ computed from the integrated fluctuations and the known temperature using the equipartition theorem \cite{Hutter1993jul}.
The cantilever frequency $f_{\text{c}}$ and ringdown time $\tau_{\text{c}}$ were determined by exciting the cantilever at resonance and measuring the decay of the induced cantilever oscillation; the cantilever quality factor was computed using $Q = \pi \: \tau_{\text{c}} \: f_{\text{c}}$.
A dissipation constant was calculated using $\Gamma = k_{\text{c}}/(2 \pi \, f_\text{c} \, Q)$ and a power spectral density of thermomechanical force fluctuations computed using $P_{\delta F} = 4 \, \Gamma \, k_{\text{b}} \, T_0$.

At a temperature of $T_0 = 4.2 \: \kelvin$, a pressure of $P = 5 \times 10^{-6} \: \milli\text{bar}$, and an applied magnetic field $B_0$ of zero, the cantilever had an apparent spring constant of $k_{\text{c}} = 1.0 \: \milli\newton \: \meter^{-1}$, a resonance frequency $f_{\text{c}} = 3500 \: \hertz$, and a quality factor of $Q = 5 \times 10^{4}$.
With the magnetic field applied parallel to the long axis of the cantilever, the cantilever's quality factor decreased to $Q = 1.0 \times 10^{4}$ at $B_0 = 1 \: \tesla$ and to $1.6 \times 10^{3}$ at $B_0 = 6 \: \tesla$.
The cantilever was positioned over the center of the coplanar waveguide in the ``hang-down'' geometry (Fig.~\ref{Fig:experimental_setup}); the cantilever was brought near the sample surface with its long axis parallel to the surface normal $\hat{z}$, an external magnetic field was applied along the $z$ direction, and the cantilever oscillated in the $x$ direction.
At $T_0 = 4.2 \: \kelvin$, over the copper centerline of the coplanar waveguide, with a tip-sample separation of $h = 1500 \: \nano\meter$, the power spectral density of cantilever force fluctuations ranged from $P_{\delta F} = 1300 \: \atto\newton^{2} \: \hertz^{-1}$ at $B_0 = 1 \: \tesla$ to $P_{\delta F} = 8100 \: \atto\newton^{2} \: \hertz^{-1}$ at $B_0 = 6 \: \tesla$.

To continuously measure the cantilever frequency, the cantilever was driven into self oscillation by making it part of an analog positive feedback loop\cite{Albrecht1991jan}, as follows: the cantilever displacement signal was measured, phase shifted by ninety degrees, amplitude limited, and fed to a piezoelectric element located below the cantilever mount.
The feedback gain was adjusted to achieve a zero-to-peak cantilever amplitude of $x_{0\text{p}} = 100 \: \nano\meter$ during the magnetic resonance experiments described below.
The cantilever displacement-versus-time signal was digitized and the instantaneous cantilever frequency determined using a software frequency demodulator \cite{Dwyer2015jan}.

\subsection{Coplanar waveguide.}

Poggio and coworkers showed that nuclear spin transitions in an MRFM experiment could be excited efficiently with a transverse magnetic field produced by passing a radiofrequency (rf) current through a microwire\cite{Poggio2007jun}.
By integrating the microwire into a coplanar waveguide \cite{Wen1969dec} we are able excite the sample at frequencies up to $20 \: \giga\hertz$ for ESR experiments.
The long axis of our waveguide's center line was oriented parallel to the $y$ axis in Fig.~\ref{Fig:experimental_setup}.

The coplanar waveguide (CPW) was fabricated in two sections.
The first section served to couple microwaves (MW) from a semi-rigid coaxial cable with an SMA connector to a CPW, made of copper plated onto an Arlon substrate.
The second section consisted of a copper CPW microfabricated on a high resistivity silicon substrate; this CPW tapered down to a 500 $\mu$m long, 10 $\mu$m wide, and 0.2 $\mu$m thick copper wire flanked on either side by ground plane.
The two sections were brought to within approximately $200 \: \micro\meter$ of each other and their center lines and ground planes were connected via multiple wire bonds.
Transmission losses were low at frequencies $\leq 5 \: \giga\hertz$ and at certain frequencies between $5$ and $20 \: \giga\hertz$--- presumably line resonances.
For electron spin resonance experiments, the irradiation frequency was set to one of these line resonances.

\subsection{Probe and nanopositioning.}

Experiments were performed at $T_0 = 4.2 \: \kelvin$, nominal $P = 5 \times$ 10$^{-6} \: \milli\text{bar}$, and at external fields from $B_0 = 0.6$ to $6 \: \tesla$ using a custom-built magnetic resonance force microscope.
The CPW and sample were affixed to a stationary cooling block while the cantilever and associated driving piezo and optical fiber were mounted on a custom-built scanner.
Coarse $x$, $y$ and $z$ scanning was achieved using custom-built Pan-style walkers, \cite{Pan1993sep} while fine motion was achieved with a piezo-tube actuator.
Three fiber-optic interferometers were used to observe the motion of the cantilever holder relative to the sample and CPW.
The Pan walkers were used to position the cantilever over the centerline of the CPW.
Alignment of the cantilever to the CPW was registered by observing a small shift in the cantilever frequency fortuitously present when the cantilever was located at the edge of the CPW's center line or ground plane.
The cantilever was brought into contact with the sample surface using a combination of coarse and fine motion; the $h=0$ location was determined by gently touching the cantilever to the surface while looking for the cantilever to stop oscillating and undergo a small buckling motion.

\subsection{Spin detection and modulation.}

Magnetic resonance signals from both nuclear spins and electron spins were observed using the force-gradient detection protocol CERMIT (Cantilever-Enabled Readout of Magnetization Inversion Transients) \cite{Garner2004jun,Mamin2007may,Moore2009dec,Vinante2011dec,Alexson2012jul,Chen2013apr,Chen2013jul}.

\subsubsection{Nuclear magnetic resonance.}
\label{Sec:MM-NMR}

Nuclear magnetic resonance (NMR) signal from $^{1}\text{H}$ Curie-law  magnetization was detected at fields between 4 and 6 tesla.
Frequency-modulated sine and cosine waves were generated at $27 \: \mega\hertz$ using a National Instruments PXI-5421 arbitrary waveform generator and up-converted to a final rf frequency $f_{\text{rf}}$ between $170$ and $260 \: \mega\hertz$ using single-sideband mixing.
The resulting frequency-modulated radiowaves were amplified at room temperature (Kalumus Model 320CP-CE) and delivered to the CPW at $4.2 \: \kelvin$ through a combination of flexible and semi-rigid coaxial cables equipped with SMA connectors.
An adiabatic rapid passage (ARP) through resonance (linear sweep; width $\Delta f_{\text{rf}} = 1 \: \mega\hertz$, except where noted) was used to invert $^{1}\text{H}$ magnetization.
Each passage lasted between one and ten cantilever cycles ($T_{\text{c}}$) and was triggered to start when the cantilever was at maximum displacement.
The FM-modulated rf inverted the sample's $^{1}\text{H}$ magnetization in a region of the sample --- a ``resonant slice'' --- whose location and size was determined by the static field $B_0$, tip field $B_{\text{tip}}$ (e.g., tip-sample separation $h$), rf center frequency $f_{\text{rf}}$, and $\Delta f_{\text{rf}}$.
The inversion of the sample's Curie-law $^{1}\text{H}$ magnetization was detected as a dc shift of the cantilever's resonance frequency\cite{Garner2004jun}.

\subsubsection{Electron spin resonance.}
\label{Sec:MM-ESR}

Following Moore \emph{et al.}\ \cite{Moore2009dec}, electron spin resonance (ESR) from Curie-law electron-spin magnetization was detected near $0.6 \: \tesla$.
Amplitude-modulated 18.5 GHz microwave irradiation (Anritsu-Wiltron source, model 6814B; American Microwave Corporation switch, model SWN-218-2DT, options 912 and B05HS20NS; Narda Microwave amplifier, model DBP-0618N830) was delivered to the CPW through a second coaxial cable.
Microwave delivery was timed to start when the cantilever was at its maximum displacement.
Microwave irradiation was applied for one cantilever cycle, during which time the cantilever motion swept out a region of partially saturated electron spin magnetization in the sample.
A $1 \, T_{\text {c}}$ interval of irradiation was followed by a $2 \, T_{\text {c}}$ interval during which no MW irradiation was applied to avoid sample heating.
As in Ref.~\citenum{Moore2009dec}, this on/off modulation  sequence was interspersed with intervals of no irradiation in order to impose a square-wave modulation on the spin-induced cantilever frequency shift.
The modulation frequency $f_{\text{mod}}$ was set to between 4 and 20 Hz to avoid $1/f$ frequency noise from sample dielectric fluctuations and $\propto f^2$ frequency noise arising from white voltage noise in the interferometer circuitry \cite{Yazdanian2008jun}.
With $f_{\text{mod}}$ so chosen, the cantilever frequency noise in the ESR experiment was close to the thermomechanical limit.
The spin-induced frequency shift was obtained from the frequency-demodulator output using a software lock-in detector.

The electron spin magnetization was measured for various microwave powers $P$ and fit to the following equation to obtain a value for the coil constant $c_{\text{p}}$ of the coplanar waveguide:
\begin{equation}
\delta f_{\text{c}}
	= \delta f_{\text{c}}^{\, \text{peak}} \:
		\frac{S}{1 + S}
	\label{Eq:saturationcurve}
\end{equation}
where $\delta f_{\text{c}}^{\, \text{peak}}$ is the maximum cantilever frequency shift and the saturation parameter $S$ is given by
\begin{equation}
S = P \, c_{\text{p}}^2 \, \gamma_{\text{e}}^2 \, T_1 \, T_2
	\label{Eq:saturation}
\end{equation}
with $\gamma_{\text{e}} = 28 \: \giga\hertz \: \tesla^{-1}$ the gyromagnetic ratio of the electron, $T_1$ the electron spin-lattice relaxation time, and $T_2$ the echo decay time. The coil constant was determined assuming \cite{Moore2009dec} $T_1 = 1.3 \: \milli\second$ and $T_2 =  450 \: \nano\second$.

\subsection{Dynamic nuclear polarization.}
\label{Sec:MM-DNP}

Dynamic nuclear polarization experiments were performed at 0.6~tesla.
Microwaves and radiofrequency waves were applied to the CPW simultaneously through two separate SMA connections on either end of the waveguide.
 An rf isolator was used to keep transmitted rf from damaging the microwave amplifier and a low-pass filter and attenuator eliminated transmitted microwaves before reaching the rf amplifier.
The cantilever's resonance frequency was recorded continuously during each DNP experiment.
To create DNP-enhanced nuclear magnetization, electron spins in the sample were saturated by applying microwave irradiation starting at time $t = 0$; while on, the microwaves were continuously modulated in a $T_{\text{c}}$-on: $2 \, T_{\text{c}}$-off sequence.
Subsequently, the cantilever frequency shifted because of microwave heating, saturation of electron-spin magnetization, and buildup of nuclear-spin magnetization.
To infer the change in cantilever frequency arising from nuclear magnetization, with the microwaves still on a single ARP was applied at time $t = \tau$ to invert the $^{1}\text{H}$ magnetization as described previously in Sec.~\ref{Sec:MM-NMR}.
The resulting frequency shift $\delta f_{\text{c}}$ was fit to
\begin{equation}
\delta f_{\text{c}}(\tau)
	= \delta f_{\text{c}}^{\, \text{max}} \:
	(1 - e^{\, -\tau/\tau_{\text{buildup}}})
	\label{Eq:buildup}
\end{equation}
where $\delta f_{\text{c}}^{\, \text{max}}$ is the maximum, steady-state frequency shift due to microwave-enhanced $^{1}\text{H}$ magnetization and $\tau_{\text{buildup}}$ is the time constant associated with the enhancement.

Real-time measurements of the spin relaxation time were performed following Alexson and coworkers \cite{Alexson2012jul}.
After an interval of DNP, an ARP was applied to invert the $^{1}\text{H}$ magnetization.
With the microwaves still on, the cantilever frequency shift was recorded with a commercial frequency counter (Stanford SR620) as a function of time $t$ and fit to
\begin{equation}
\delta f_{\text{c}}(t)
	= \delta f_{\text{c}}^{\, \text{initial}} \:
	e^{\, -t/T_1^{\text{eff}}}
	\label{Eq:decay}
\end{equation}
to obtain $\delta f_{\text{c}}^{\, \text{initial}}$, the initial frequency shift due to microwave-enhanced $^{1}\text{H}$ magnetization, and $T_1^{\text{eff}}$, an effective $^{1}\text{H}$ spin-lattice relaxation time with the microwaves on.

\begin{figure*}[th!]
\centering
\includegraphics[width=\textwidth]{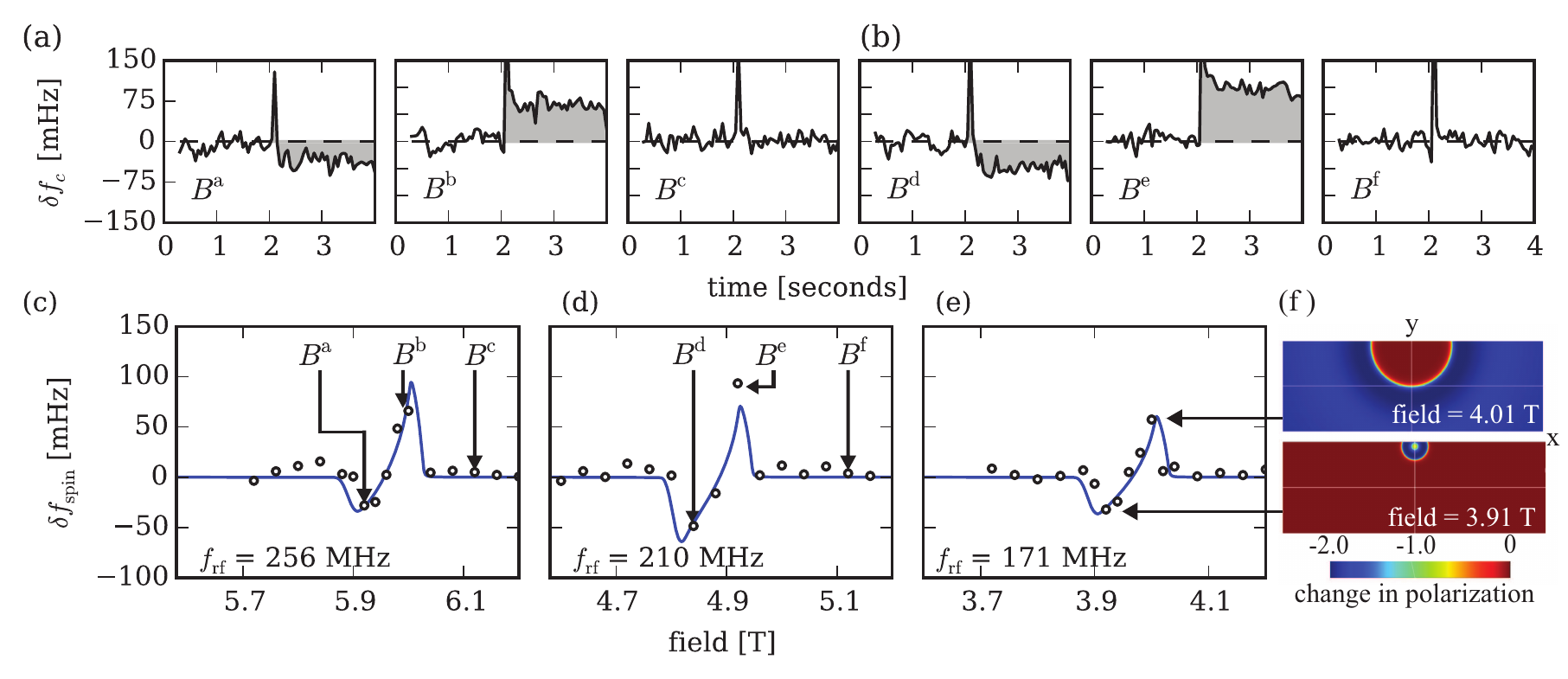}
\caption{Force-gradient-detected nuclear magnetic resonance signal. Upper: Cantilever resonance frequency shift $\delta f_{\text{c}}$ versus time at selected values of the rf center frequency $f_{\text{rf}}$ and magnetic field $B_0 = B^{\text{a}}, B^{\text{b}}, \ldots, B^{\text{f}}$.
A single adiabatic rapid passage through resonance was applied at time $t = 2 \: \second$ to invert $^{1}\text{H}$ nuclear spin magnetization.
Center frequency: (a) $256 \: \mega\hertz$ and (b) $210 \: \mega\hertz$.
Bottom: Observed (open black circles) and calculated (solid blue line) rf-induced cantilever frequency shift --- the spin signal, $\delta f_{\text{spin}}$ --- versus magnetic field for three experiments centered at fields of (c) $6.0 \: \tesla$, (d) $5.0 \: \tesla$, and (e) $4.0 \: \tesla$.
The rf center frequency $f_{\text{rf}}$ is indicated in the lower left of each subfigure (c,d,e).
(f) Calculated resonant slice of magnetization at $f_{\text{rf}} = 171 \: \mega\hertz$ and two selected fields.
The color indicates the change in spin polarization induced by the adiabatic rapid passage.
Upper: at $4.01 \: \tesla$, ``bulk'' spins far away from the magnetic tip are inverted.
Lower: at $3.91 \: \tesla$, ``local'' spins right below the tip are inverted.
Experimental parameters: tip-sample separation $h = 1500 \: \nano\meter$ and rf frequency sweep width $\Delta f_{\text{rf}} = 1 \: \mega\hertz$.
\label{Fig:1H_NMR_various_fields}}
\end{figure*}

\subsection{Signal simulation.}

The cantilever frequency shift was calculated using Eq.~\ref{Eq:freq_shift} and measured values for $f_{\text{c}}$ and k$_{\text{c}}$.
The sum over spins in resonance in Eq.~\ref{Eq:freq_shift} was evaluated numerically, approximating the sample as a collection of independent spin $1/2$ particles.
Equation~\ref{Eq:freq_shift} is valid when the cantilever amplitude, $x_{\text{0p}} = 0.1 \: \micro\meter$, is small compared to the distance $r + h = 5.0 \: \micro\meter$ between the center of the magnetic sphere and the sample, which is the case here.

The tip magnetic field component $B_{z}^{\text{tip}}$ and second derivative $G_{zxx}^{\text{tip}}$ were calculated using analytical formulas for a uniformly magnetized sphere.
The sample was modeled as a box having dimensions $\Delta x = 10 \: \micro\meter$ (the width of the CPW center line), $\Delta y = 30 \: \micro\meter$, and $\Delta z = 0.25 \: \micro\meter$.
At each point in the sample, the change in electron or nuclear Curie-law magnetization $\Delta \mu_{z,k}$ due to either inversion or saturation was computed using the Bloch equations.

\subsubsection{Nuclear magnetic resonance.}

The sample box was approximated using $N_x = 250$, $N_y = 750$, and $N_z = 13$ grid points.
The simulations employed a proton density of $\rho_{\text{p}} = 49 \: \text{spins} \: \nano\meter^{-3}$ and a proton magnetic moment of $\mu_{\text{p}} = 1.4106 \times 10^{-26} \: \joule \: \tesla^{-1}$.
The magnitude of the (assumed perfectly homogeneous) transverse oscillating magnetic field was taken to be $B_1 = 2.5 \: \text{mT}$.
This number was obtained from electromagnetic simulations of the CPW (Sonnet Software, Inc.) carried out using the measured input power of $200 \: \milli\watt$.
The Bloch equations were used to compute $\Delta \mu_{z,k}$ under the assumption that spin-locked magnetization followed the effective field in the rotating frame adiabatically.
The only free parameter in each simulation was a lateral $x$ offset between the center of the magnetic sphere and the center of the CPW; this parameter was adjusted to achieve improved agreement between the measured and calculated frequency shift versus magnetic field curves.

\subsubsection{Electron spin resonance.}

The sample box was approximated using $N_x = 400$, $N_y = 1200$, and $N_z = 13$ grid points.  It was assumed that the magnetic sphere was positioned directly over the center of the CPW.
The simulations employed an electron-spin density of $\rho_{\text{e}} = 2.41 \times 10^{-2} \: \text{spins} \: \nano\meter^{-3}$, an electron magnetic moment of $\mu_{\text{e}} = -928.4 \times 10^{-26} \: \joule \: \tesla^{-1}$, $T_1 = 1.3 \: \milli\second$, $T_2 = 450 \: \nano\second$, and $B_1 = 1.3 \: \micro\tesla$.
The $B_1$ was taken from a Sonnet simulation of the CPW operating at $18.5 \: \giga\hertz$; the input power used in the simulation was computed from the estimated experimental input power and the measured transmission losses.

\section{Results}

\subsection{Nuclear magnetic resonance at high field.}

Nuclear magnetic resonance signal is shown in Fig.~\ref{Fig:1H_NMR_various_fields}.
In Fig.~\ref{Fig:1H_NMR_various_fields}(a,b) we show representative plots of cantilever frequency shift versus time acquired at six different combinations of rf center frequency and magnetic field.
The spin-inverting ARP sweep was applied at $t = 2.0 \: \second$ in each plot.
The sweep induced a transient positive frequency shift whether or not the rf was in resonance with sample spins.
When the sweep was in resonance with sample spins it induced a long-lived shift in the cantilever frequency.   This long-lived shift --- the mechanically-detected nuclear spin signal --- is highlighted in gray in Fig.~\ref{Fig:1H_NMR_various_fields}(a,b).
A spin signal was calculated by subtracting the cantilever frequency before and after the ARP sweep, with the ``after'' time window adjusted to reject the spurious rf-induced frequency-shift transient.

In Fig.~\ref{Fig:1H_NMR_various_fields}(c,d,e) we plot the resulting spin signal acquired as a function magnetic field $B_0$ at three different rf center frequencies.
The spin signal is either positive or negative, depending on the exact rf center frequency and field \cite{Garner2004jun}.
In Fig.~\ref{Fig:1H_NMR_various_fields}(f) we show the calculated spatial distribution of the change in spin polarization induced by the ARP at two $B_0$ values in the experiment of Fig.~\ref{Fig:1H_NMR_various_fields}(e).
At $4.01 \: \tesla$, the spin signal is dominated by ``bulk spins'' far away from the magnet where $G_{zxx}^{\text{tip}}$ is positive.
At $3.91 \: \tesla$, in contrast, the spin signal is dominated by ``local spins'' close to the magnet where $G_{zxx}^{\text{tip}}$ is negative.
The calculated signal is shown as solid lines in Fig.~\ref{Fig:1H_NMR_various_fields}(c,d,e).
Both the absolute size of the nuclear spin signal and its complicated dependence on field are in excellent agreement with simulations and consistent with our observing Curie-law magnetization from $^{1}\text{H}$ spins at $T_0 = 4.2 \, \kelvin$.

\subsection{Electron-spin resonance at low field.}

\begin{figure}
\centering
\includegraphics[width=3.50in]{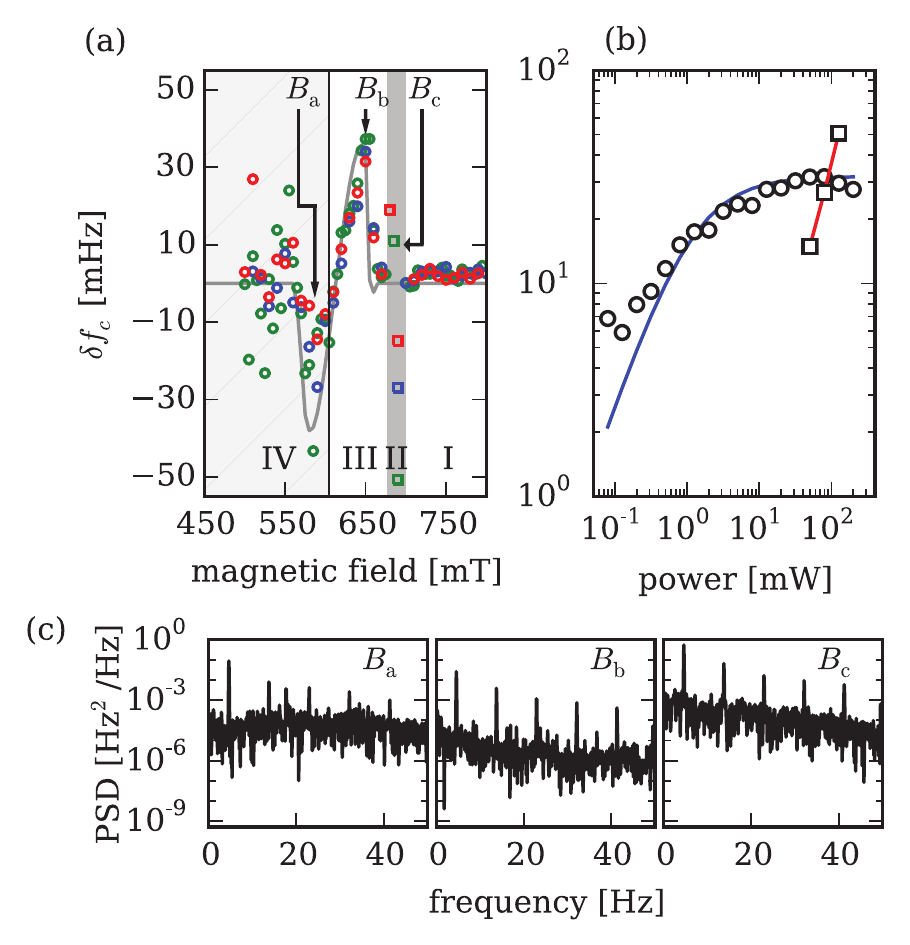}
\caption{Force-gradient-detected electron-spin resonance signal.
(a) Observed (circles) and calculated (grey line) spin-induced cantilever frequency shift versus magnetic field at three different microwave powers: $50 \: \milli\watt$ (red circles), $79 \: \milli\watt$ (blue circles), and $126 \: \milli\watt$ (green circles).
(b) Absolute value of the cantilever frequency shift in $\milli\hertz$ versus microwave power with the field set to be in resonance (field $B_0 = B_{\text{b}}$; circles) and out of resonance ($B_0 = B_{\text{c}}$; squares) with electron spins.
The solid blue line is a fit to Eqs.~\ref{Eq:saturationcurve} and \ref{Eq:saturation} to give a coil constant $c_{\text{p}} = 1.6 \, \mu T/ \sqrt{\text{mW}}$ at 18.5 GHz; the red line is a guide to the eye.  Note the log-log scale.
(c) Cantilever frequency-shift power spectrum for the field set to be on resonance (center) and off resonance (left and  right).
Note the logarithmic $y$ axis.
A modulated cantilever-frequency signal is apparent in each plot as a large peak at $f = f_{\text{mod}}$ and its harmonics.
The noise floor is orders of magnitude higher with the field set off resonance to $B_{\text{a}}$ and $B_{\text{c}}$.
Experimental parameters: tip-sample separation $h = 1500 \: \nano\meter$ and microwave frequency $f_{\text{MW}} = 18.5 \: \giga\hertz$.
\label{Fig:ESR_vs_power-saturation}}
\end{figure}

The applied magnetic field was reduced to near $0.6 \: \tesla$ and electron-spin resonance signal was acquired as a function of magnetic field as described in the Methods section (see Fig.~\ref{Fig:ESR_vs_power-saturation}).
The observed frequency shift versus magnetic field had a lineshape similar to that seen in the NMR case but with larger deviations from the calculated signal.

No signal was observed at high field (Fig.~\ref{Fig:ESR_vs_power-saturation}(a); region I), as expected.
As the field was lowered, large microwave-induced shifts in cantilever frequency were seen at certain magnetic fields (Fig.~\ref{Fig:ESR_vs_power-saturation}(a); grey shaded region II).
The magnitude of these signals depended linearly on microwave power (Fig.~\ref{Fig:ESR_vs_power-saturation}(b); red line) and their associated frequency-shift power spectra exhibited large, low-frequency fluctuations (Fig.~\ref{Fig:ESR_vs_power-saturation}(c); right); we tentatively attribute the region II frequency shifts to spurious excitation of ferromagnetic resonances in the cantilever tip.
As the field was lowered further, sample spins came into resonance and we observed the expected spin-induced changes in cantilever frequency.
The size and lineshape of the region-III signal agreed well with the simulated force-gradient ESR signal (Fig.~\ref{Fig:ESR_vs_power-saturation}(a); grey line).
The region-III frequency shifts had a dependence on microwave power consistent with saturation of an electron-spin magnetic resonance signal (Fig.~\ref{Fig:ESR_vs_power-saturation}(b); blue line).
When the field was lowered further, to below 580 mT (region IV; Fig.~\ref{Fig:ESR_vs_power-saturation}(a)), large variations in the cantilever frequency shift versus field signal were observed.
There variations arose, we hypothesize, from either ferromagnetic resonances or magnetization fluctuations in the cantilever's magnetic tip; below $B_0 = 580 \: \milli\tesla$, the magnetization may not be fully saturated at all locations in the magnet given that the saturation magnetization for nickel is $\mu_{\text{0}} \, \text{M} = 600 \: \milli\tesla$.

In summary, while spurious frequency-shift signals were present in regions II and IV of the Fig.~\ref{Fig:ESR_vs_power-saturation} electron-spin resonance signal, a microwave-induced spin signal could be observed in field region III whose dependence on magnetic field and microwave power was consistent with magnetic resonance signal from Curie-law electron spin magnetization at $T_0 = 4.2 \: \kelvin$.

\subsection{Microwave-enhanced nuclear magnetic resonance at low field.}

\begin{figure}
\centering
\includegraphics[width=3.25in]{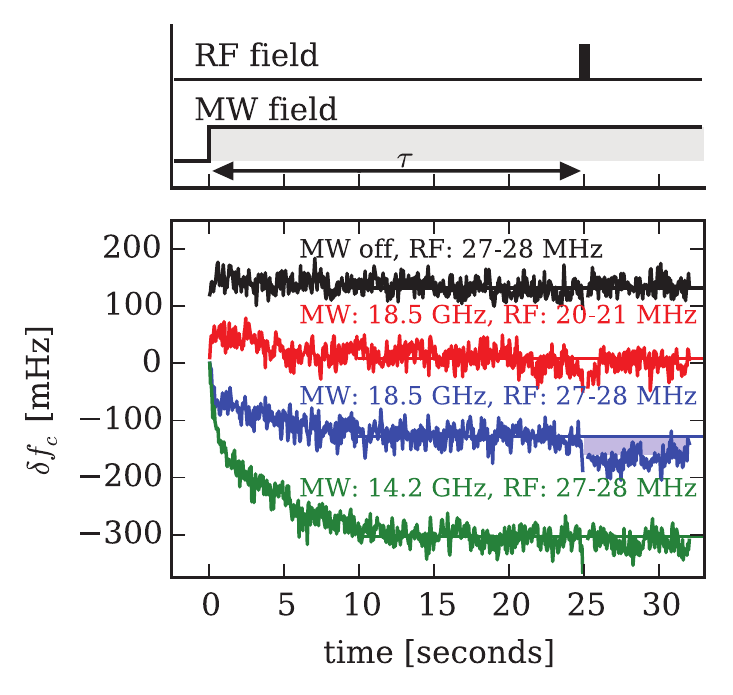}
\caption{Evidence for dynamic nuclear polarization in a magnetic resonance force microscope experiment.
The experiment was carried out on a TEMPAMINE-doped polystyrene sample at $B_0 = 655 \: \milli\tesla$ and $T_0 = 4.2 \: \kelvin$.
Upper: Microwaves were turned on at time $t = 0 \: \second$ to saturate electron spins (Section~\ref{Sec:MM-DNP}).
A duration ${\Delta t}_{\text{rf}} = 10 \: T_{\text{c}}$ adiabatic rapid passage through resonance was applied at time $\tau = 25 \: \second$ to invert nuclear spins.
Lower: Cantilever frequency shift versus time.
From top to bottom: microwaves absent, rf on resonance (black line); microwaves on resonance, rf off resonance (red line); microwaves on resonance, rf on resonance (blue line); and microwaves off resonance, rf on resonance (green line).
The traces have been offset vertically for clarity.
\label{Fig:DNP_control_expts}}
\end{figure}

The above experiments demonstrate our ability to excite nuclear spins at high field and electron spins at low field using a single coplanar waveguide.
We detected magnetic resonance in both experiments as a change in the mechanical resonance frequency of a magnet-tipped cantilever.
We next looked for evidence that hyperthermal nuclear magnetization could be created in our microscope at low field via the dynamic nuclear polarization effect.
We emphasize that the same sample, waveguide, and magnet-tipped cantilever was employed in all three experiments.

Microwaves and radiowaves were applied as indicated in the timing diagram of Fig.~\ref{Fig:DNP_control_expts}.
To demonstrate DNP, cantilever frequency was recorded as a function of time in experiments employing combinations of on- and off-resonance microwaves and rf.
A frequency shift due to nuclear magnetization was observed following a period of on-resonance microwave irradiation.
This microwave-enhanced nuclear-spin signal is shown shaded in blue in Fig.~\ref{Fig:DNP_control_expts}.
No such nuclear-spin signal was observed when either the microwaves or the radiowaves were applied off resonance.
Application of microwaves led to a decrease in cantilever frequency at times $0 < t \leq 15 \: \second$; because it was present when both on- and off-resonance microwave irradiation was applied, we attribute this decrease to a heating-related artifact.
Significantly, no nuclear spin signal could be detected at $B_0 = 0.6 \: \tesla$ without first applying microwaves.
This is expected.
The Curie-law nuclear-spin signal apparent at $B_0 = 6 \: \tesla$ in Fig.~\ref{Fig:1H_NMR_various_fields} scales linearly with $B_0$; we would predict the Curie-law signal at $B_0 = 0.6 \: \tesla$ to fall below the cantilever frequency noise floor and therefore be undetectable.
Comparing the nuclear spin signal observed following microwave irradiation at $B_0 = 0.6 \: \tesla$ to the  nuclear spin signal extrapolated from the $B_0 = 6 \: \tesla$ experiments, we can nevertheless estimate that the applied microwaves are inducing a hyperthermal nuclear magnetization enhanced by a factor of between $\epsilon = 10$ and $20$.

\begin{figure*}[t!]
\centering
\includegraphics[width=6.50in]{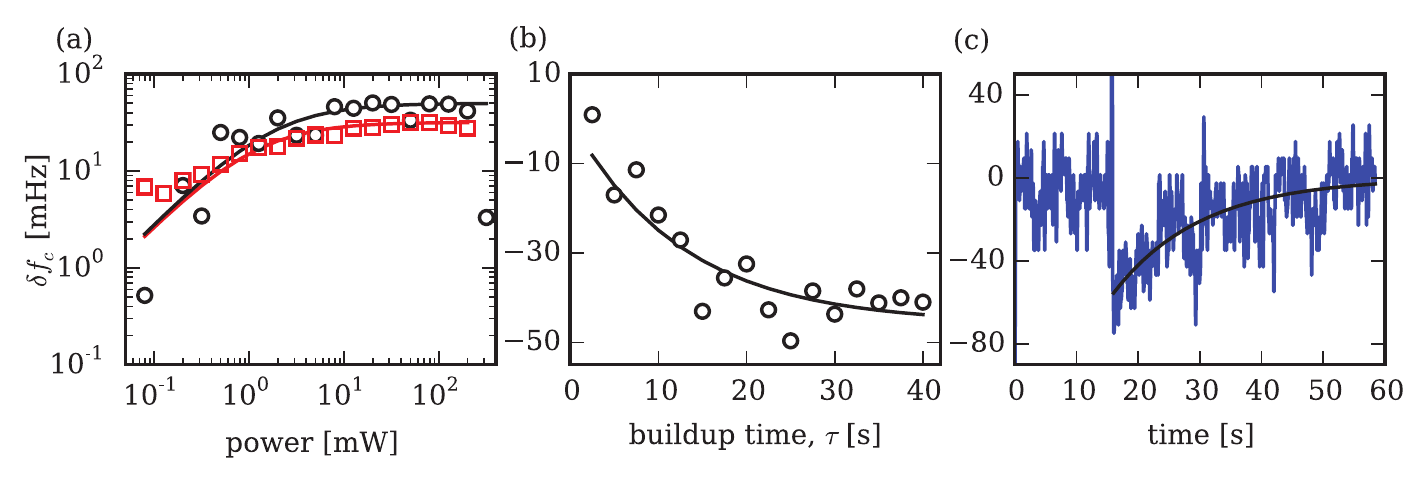}
\caption{Dependence of the microwave-enhanced nuclear spin signal on microwave power and time.
(a) Absolute value of the nuclear-spin frequency shift following a single ARP (black circles) versus microwave power and, for comparison, electron-spin signal (red squares) versus microwave power. The solid red line is a fit to Eqs.~\ref{Eq:saturationcurve} and \ref{Eq:saturation}. The black line is a guide to the eye.
(b) Nuclear-spin frequency shift acquired after $\tau$ seconds of microwave irradiation.
The solid line is a fit to Eq.~\ref{Eq:buildup}.
(c) Decay of the nuclear-spin frequency shift.
A single ARP was applied at time $t = 15 \: \second$ to invert nuclear spin magnetization; microwaves were applied in an on/off pattern continuously.
The solid line is a fit of the $t \geq 15 \: \second$ data to Eq.~\ref{Eq:decay}.
Experimental parameters: $h = 1500 \: \nano\meter$,
$f_{\text{MW}} = 18.5 \: \giga\hertz$, and $B_0 = 0.655 \: \tesla$; in (b), the ARP had an initial and final frequency of $f_{\text{rf}}^{\text{initial}} = 27.0 \: \mega\hertz$ and $f_{\text{rf}}^{\text{final}} = 28.0 \: \mega\hertz$, respectively; in (c), $f_{\text{rf}}^{\text{initial}} = 27.6 \: \mega\hertz$ and $f_{\text{rf}}^{\text{final}} = 28.1 \: \mega\hertz$.
\label{Fig:DNP_saturation_buildup_T1}}
\end{figure*}	

The dependence of the microwave-enhanced nuclear spin signal on microwave power is shown in Fig.~\ref{Fig:DNP_saturation_buildup_T1}(a).
Both the nuclear-spin signal and the electron-spin signal plateau at the same microwave power, suggesting that the Fig.~\ref{Fig:DNP_control_expts} nuclear-spin signal originates in the electron-spin magnetization.
The dependence of the nuclear-spin signal on microwave irradiation time is shown in Fig.~\ref{Fig:DNP_saturation_buildup_T1}(b); the measured buildup time is $\tau_{\text{buildup}} = 12.7 \pm 3.3 \: \second$.
Ideally the DNP buildup time should be compared to the nuclear spin's spin-lattice relaxation time, $T_1$.
  The $^{1}\text{H}$ spin-lattice relaxation time in the absence of microwave irradiation was difficult to measure at $B_0 = 0.6 \: \tesla$ because of the large transient change in cantilever frequency created by turning off the microwave irradiation.
An effective $^{1}\text{H}$ $T_1$ could be measured in the presence of resonant microwave irradiation by observing the cantilever frequency shift in real time \cite{Vinante2011dec,Alexson2012jul} after the application of a single ARP sweep (Fig.~\ref{Fig:DNP_saturation_buildup_T1}(c)).
The resulting cantilever frequency transient was well described by a single exponential decay having an effective spin-lattice relaxation time of $T_{1}^{\text{eff}}= 14.3 \pm 1.0 \: \second$.
This value is in reasonable agreement with the $^{1}\text{H}$ spin-lattice relaxation time $T_{1} = 30.8 \pm 0.9 \: \second$ measured at $B_0 = 5 \: \tesla$  (Supporting Information).
The factor-of-two agreement between these estimated $^{1}\text{H}$ spin-lattice relaxation times and $\tau_{\text{buildup}}$ is consistent with buildup of nuclear magnetization via a DNP effect.

\begin{figure*}[t!]
\centering
\includegraphics[width=6.00in]{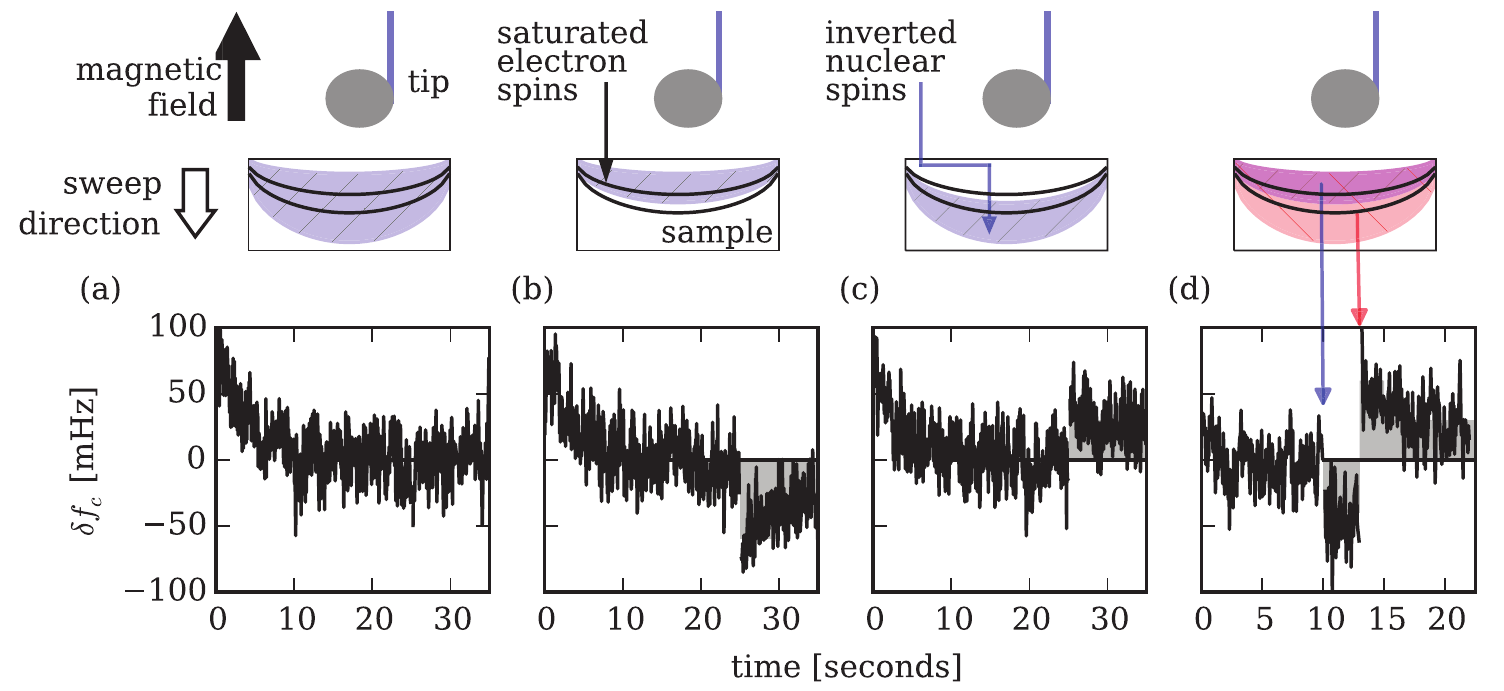}
\caption{Evidence that the \emph{net} microwave-enhanced nuclear polarization is zero.
Microwave irradiation was turned on at $t = 0 \: \second$.
(a-c) At time $t = 25 \: \second$, an ARP rf sweep was initiated that ran from (a) $27.6$ to $28.6 \: \mega\hertz$, (b) $27.6$ to $28.0 \: \mega\hertz$, and (c) $28.1$ to $28.4 \: \mega\hertz$.
(d) Two ARP sweeps were initiated: (blue) one at time $t = 10 \: \second$ running from $27.6$ to $28.1 \: \mega\hertz$ and (red) a second one at time $t = 13 \: \second$ running from $27.6$ to $28.6 \: \mega\hertz$.
The nuclear-spin induced frequency shift is shaded grey.
\label{Fig:DNP_vary_sweep_width}}
\end{figure*}	

The reader will have noticed that, in the experiments of Fig.~\ref{Fig:DNP_control_expts} and\ \ref{Fig:DNP_saturation_buildup_T1}, the rf was not exactly in resonance with nuclear spins in the center of the resonant slice defined by the applied microwaves.\footnote{This would require an ARP sweep centered at $18.50 \: \giga\hertz \times \gamma_{\text{p}}/\gamma_{\text{e}} = 28.0 \: \mega\hertz$.} If the ARP was adjusted to flip nuclear spins in a region exactly centered on the resonant slice, Fig.~\ref{Fig:DNP_vary_sweep_width}(a), then there was no observable frequency shift, implying a \emph{net} nuclear-spin enhancement of \emph{zero}.  Further experiments were carried out in which resonant microwave irradiation was applied at $t = 0 \: \second$ to initiate DNP, one or two ARP sweeps were applied subsequently to generate a frequency shift proportional to the nuclear polarization, and the center frequency of the ARP sweep was varied to selectively invert nuclear spins that were nearer to or further from the magnet tip than were the resonant electron spins.  The results of these experiments are shown in Fig.~\ref{Fig:DNP_vary_sweep_width}(b-d).  These results are consistent with the nuclear spins proximal to the magnet tip having a positive $\epsilon$ and distal spins having a negative $\epsilon$.

\begin{figure}
\centering
\includegraphics[width=3.10in]{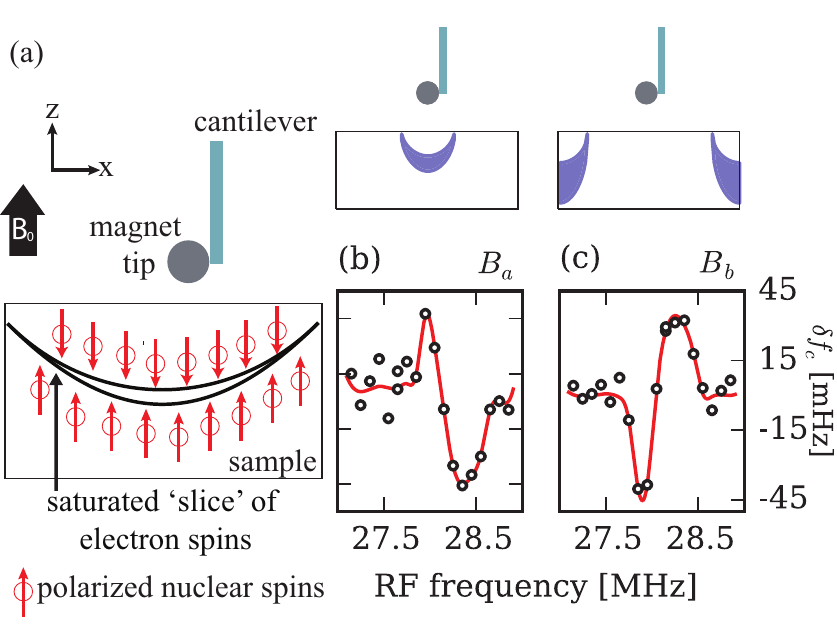}
\caption{Mapping the spatial distribution of enhanced nuclear polarization.
(a) The spatial distribution of microwave-enhanced nuclear-spin polarization consistent with the Fig.~\ref{Fig:DNP_vary_sweep_width} data.
ARP-induced change in cantilever frequency versus radiofrequency resonance offset at (b) $B_0 = 568 \: \milli\tesla$, where ``local'' electron spins are in resonance and at (c) $B_0 = 655 \: \milli\tesla$, where ``bulk'' electron spins are in resonance.
The solid red lines are guides to the eye.
The labels $B_{\text{a}}$ and $B_{\text{b}}$ refer to fields indicated in Fig.~\ref{Fig:ESR_vs_power-saturation}(a).
Upper: The corresponding regions of electron-spin magnetization in resonance.
Experimental parameters: microwave frequency $f_{\text{MW}} = 18.5 \: \giga\hertz$, ARP sweep width ${\Delta f}_{\text{rf}} = 0.30 \: \mega\hertz$, ARP sweep duration ${\Delta t}_{\text{rf}} = 1 \: T_{\text{c}}$, and rf frequency step $ = 0.10 \: \mega\hertz$.
\label{Fig:DNP_spins_sketch2}}
\end{figure}

The associated spatial distribution of enhanced nuclear magnetization is sketched in Fig.~\ref{Fig:DNP_spins_sketch2}(a).
To map out this distribution, the ARP sweep width was reduced to ${\Delta f}_{\text{rf}} = 0.3 \: \mega\hertz$, the rf center frequency varied systematically about $f_{\text{rf}} = 28 \: \mega\hertz$, and the above experiments were repeated.
The $B_0$ field was varied to examine the nuclear enhancement about both a ``local'' resonant slice, Fig.~\ref{Fig:DNP_spins_sketch2}(b), and a ``bulk'' resonant slice, Fig.~\ref{Fig:DNP_spins_sketch2}(c).
Considering that $G_{zxx}^{\text{tip}}$ is negative in the local slice and positive in the bulk slice, the Fig.~\ref{Fig:DNP_spins_sketch2}(b,c) data is consistent $\epsilon > 0$ for proximal spins and $\epsilon < 0$ for distal spins in both slices.

\section{Discussion}

In summary, the dependence of the cantilever frequency shift seen in Figs.~\ref{Fig:DNP_control_expts}, \ref{Fig:DNP_saturation_buildup_T1}, \ref{Fig:DNP_vary_sweep_width}, and \ref{Fig:DNP_spins_sketch2} on
\begin{enumerate}
	{
	\setlength{\itemsep}{0pt}
    \setlength{\parskip}{0pt}
    \setlength{\parsep}{0pt}

	\item the longitudinal magnetic field;
	\item the frequency and timing of the
		applied rf; and
	\item the microwaves' frequency, timing,
		and intensity	
	}
	\end{enumerate}
is consistent with signal arising from DNP-enhanced nuclear magnetization interacting with a magnet-tipped cantilever.
A number of DNP mechanisms \cite{Maly2008feb,Prisner2008aug} are possible in TEMPAMINE-doped polystyrene, including the solid effect, thermal mixing, the cross effect, and the recently proposed separative magnetization transport (SMT) mechanism \cite{Picone2014,Picone2015jan,Sidles2015jul}.
The spatial distribution of the DNP enhancement $\epsilon$ in Fig.~\ref{Fig:DNP_spins_sketch2} is inconsistent with the SMT mechanism alone.
The relative values of line widths and resonance frequencies govern which of the established DNP mechanisms is active in a sample at a given field.
The homogeneous linewidth of the ESR spectrum in our sample is $\delta = 1/(\gamma_{\text{e}} T_{2\text{e}}) = 2 \: \mega\hertz$.  At $B_0 = 0.6 \: \tesla$, the inhomogeneous linewidth of the ESR spectrum is approximately $\Delta \sim 77 \: \mega\hertz$ (due primarily to hyperfine anisotropy, $\sim 77 \: \mega\hertz$, and not $g$-factor anisotropy, $\sim 61 \: \mega\hertz$).
The nuclear Larmor frequency is $\omega = 28 \: \mega\hertz$ here.
These line widths and frequencies satisfy \cite{Griffin2013jan} $\Delta > \omega > \delta$, making cross-effect the most likely DNP mechanism at play in our experiment.

{
\renewcommand{\arraystretch}{1.25}
\begin{table}[t]
\small
\caption{Relevant properties of proton spins and unpaired electron spins in polystyrene doped with $40 \: \milli\text{M}$ TEMPAMINE}
\label{Table:Props}
\begin{tabular*}{0.48\textwidth}{@{\extracolsep{\fill}}rlrrl}
\hline
\multirow{2}{*}{quantity}
	& &
	\multicolumn{2}{c}{value}
	& \multirow{2}{*}{unit} \\
& & electron & $^{1}\text{H}$ & \\ \hline
concentration
	& $\rho$
	& $2.41 \times 10^{-2}$ & $49$
	& $\text{spins} \: \nano\meter^{-3}$ \\
interspin spacing
	& $a$
	& $3.5$ & $0.27$
	& $\nano\meter$ \\
local field$^{\text{a,b}}$
	& $B_{\text{L}}$
	& $0.17$ & $0.19$
	& $\milli\tesla$ \\
diffusion constant$^{\text{c,d}}$
	& $D$
	& $58 \times 10^{6}$ & $450$
	& $\nano\meter^2 \: \second^{-1}$ \\
relaxation time$^{\text{e,f}}$
	& $T_1$
	& $1.3 \times 10^{-3}$ & $14.3$
	& $\second$ \\
diffusion length$^{\text{g}}$
	& $\ell_{\text{D}}$
	& 275 & 80
	& $\nano\meter$ \\
$1^{\text{st}}$ critical gradient$^{\text{h}}$
	& $G^{\,\text{crit}}_{1}$
	& $0.0006$ & $0.003$
	& $\milli\tesla \: \nano\meter^{-1}$ \\
$2^{\text{nd}}$ critical gradient$^{\text{i}}$
	& $G^{\,\text{crit}}_{2}$
	& $0.05$ & $0.7$
	& $\milli\tesla \: \nano\meter^{-1}$ \\
\hline
\end{tabular*}
\vspace{0.05in} \\
$^{\text{a}}$Calculated using $B_{\text{L}} = 7.6 \mu_0 \mu_{\text{e}} \rho / 4 \pi$, adapted from Ref.~\citenum{Budakian2004jan}, with $\rho$ the electron spin density, $\mu_{\text{e}} = 9.28 \: \atto\newton \: \nano\meter \: \milli\tesla^{-1}$ the electron magnetic moment, and $\mu_0 / \ 4 \pi = 0.1 \: \milli\tesla^2 \: \nano\meter^2 \: \atto\newton^{-1}$ the free-space permeability (in practical units).
$^{\text{b}}$Calculated as $H_{\text{L}}/\gamma_{\text{p}}$ with $H_{\text{L}} = 8 \: \kilo\hertz$ taken from Afeworki \emph{et al.}, Ref.~\citenum{Afeworki1992aug2}.
$^{\text{c}}$Estimated using $D = \gamma_e B_{\text{L}} a^2$.
$^{\text{d}}$From Afeworki \emph{et al.}, Ref.~\citenum{Afeworki1992aug2}.
$^{\text{e}}$From Ref.~\citenum{Moore2009dec}.
$^{\text{f}}$\emph{Vide supra}.
$^{\text{g}}$Calculated using $\ell_{\text{D}} = \sqrt{D T_1}$.
$^{\text{h}}$Calculated using Eq.~\ref{Eq:Gcrit1}.
$^{\text{i}}$Calculated using Eq.~\ref{Eq:Gcrit2}.
\end{table}
}	

We expect the large magnetic field gradient present in our experiment to have a number of effects on the sample's nuclear and electron spins that we will now consider.
Relevant properties of our sample's $^{1}\text{H}$ and electron spins are summarized in Table~\ref{Table:Props}.

\begin{figure}
\centering
\includegraphics[width=2.0in]{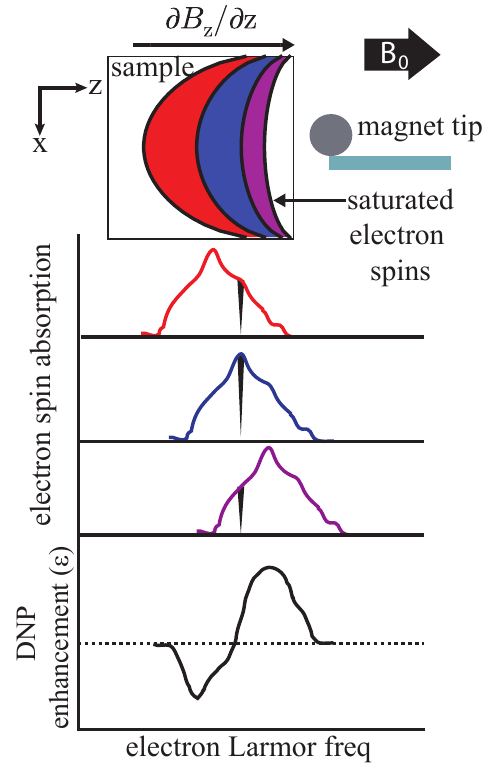}
\caption{Microwave irradiation applied to saturate electron spins in one resonant slice (blue slice; middle abs.\ curve) simultaneously saturate electron spins at (1) the high-$\nu$ end of the $a(\nu)$ profile of distal nitroxide molecules (red slice; top abs.\ curve), leading to a $\epsilon < 0$ nuclear polarization enhancement, and (2) the low-$\nu$ end of the $a(\nu)$ profile of proximal nitroxide molecules (purple slice; lower abs.\ curve), leading to a $\epsilon > 0$ enhancement. Bottom: The expected polarization profile.\label{Fig:nitroxide_saturation}}
\end{figure}

The presence of the cantilever tip's magnetic field gradient  qualitatively explains why cross-effect DNP would produce the enhancement profile shown in Fig.~\ref{Fig:DNP_spins_sketch2}.
In a homogeneous magnetic field, an ensemble of randomly oriented nitroxide molecules will absorb microwaves across a distribution of frequencies $\nu$ because of the anisotropy in the nitroxide electron's hyperfine coupling and $g$ tensor.
In our experiment, each nitroxide's resonance frequency is shifted by an additional amount inversely proportional to its distance from the cantilever's magnetic tip.
The observed polarization profile can be understood by considering both these effects together (Fig.~\ref{Fig:nitroxide_saturation}).
The observed profile is rationalized by realizing that applied microwaves simultaneously excite distal molecules at the high-$\nu$ end of the absorption profile $a(\nu)$, yielding $\epsilon < 0$ proton polarization, and proximate molecules at the low-$\nu$ end of the $a(\nu)$ profile, yielding $\epsilon > 0$ proton polarization.

The picture of Fig.~\ref{Fig:nitroxide_saturation} predicts that the nuclear spin enhancement should be limited to distances away from the central resonance slice where the electron Larmor frequency is shifted by no more than $\pm \omega$ due to the magnetic field gradient.
The electron spins directly below the magnetic tip experience a gradient of $G^{\mathrm{tip}}_{zz} = - 2 \mu_0 M_{\text{sat}} a^3 (h + a)^{-4}$.  With $\mu_0 M_{\text{sat}} = 600 \: \milli\tesla$ for nickel, a tip radius of $a = 3500 \: \nano\meter$, and a tip-sample separation of $h = 1500 \: \nano\meter$, the gradient is estimated to be $G^{\mathrm{tip}}_{zz} = 0.082 \: \milli\tesla \: \nano\meter^{-1}$.  The peak enhancement is therefore predicted to occur at a distance $\Delta z =  \omega/(\gamma_{\text{e}} G^{\mathrm{tip}}_{zz}) = 12 \: \nano\meter$ to either side of the resonant slice.  The nuclear spin polarization in Fig.~\ref{Fig:DNP_spins_sketch2}(b,c) is observed to peak at a distance approximately $0.25 \: \mega\hertz/(\gamma_{\text{p}} \:  G^{\text{tip}}_{zz}) = 72 \: \nano\meter$ from the center of the resonant slice, a distance that is good agreement with the calculated proton spin-diffusion length $\ell_{\text{D}} = \sqrt{D T_1} = 80 \: \nano\meter$; see Table~\ref{Table:Props}.  Based on the calculated $\Delta z$, we would expect spin diffusion during DNP to lead to a ``blurring'' of the enhanced nuclear magnetization in our experiment.   This blurring of the bipolar magnetization profile by spin diffusion may account for the modest 1.5 to 3.0 \% efficiency of DNP seen here.

The large magnetic field gradient present in an MRFM experiment has been shown to affect both electron \cite{Budakian2004jan} and nuclear \cite{Eberhardt2007nov} spin diffusion.
For each kind of spin, there are two effects to consider.  Eberhardt and co-workers \cite{Eberhardt2007nov} studied nuclear spin diffusion and showed that spin diffusion was impeded when the gradient was large enough to be felt by diffusing spin polarization during the spin polarization's lifetime $T_1$.
They concluded that the onset of reduced spin diffusion should occur at a critical gradient of
\begin{equation}
G^{\,\text{crit}}_{1}
	= \frac{B_{\text{L}}}{\ell_{\text{D}}}
	= \frac{B_{\text{L}}}{\sqrt{D T_1}}
	\label{Eq:Gcrit1}
\end{equation}
where $B_{\text{L}}$ (in our notation) is the mean-square ``local field'' produced by dipolar interactions with the other spins in the sample.
Budakian \emph{et al.}\ \cite{Budakian2004jan}, in their MRFM study of electron spin-lattice relaxation, argued that spin diffusion should be quenched entirely when the resonance frequency difference at different lattice sites is larger than the spin-spin dipolar coupling.
According to this argument, the onset of spin diffusion quenching should occur at a critical gradient of approximately
\begin{equation}
G^{\,\text{crit}}_{2}
	= \frac{B_{\text{L}}}{a}
	\label{Eq:Gcrit2}
\end{equation}
with $a$ the lattice spacing.

In Table~\ref{Table:Props} we provide estimates of $G^{\,\text{crit}}_{1}$ and $G^{\,\text{crit}}_{2}$ for both the $^{1}\text{H}$ and electron spins in our sample.
For electron spins, we estimate that $G^{\mathrm{tip}}_{zz} \geq G^{\,\text{crit}}_{2} > G^{\,\text{crit}}_{1}$; electron spin diffusion should be quenched or at least strongly affected by the magnetic field gradient present in our experiment.
For the $^{1}\text{H}$ spins, on the other hand, $G^{\,\text{crit}}_{2} > G^{\mathrm{tip}}_{zz} > G^{\,\text{crit}}_{1}$; we should be in a regime where nuclear spin diffusion is impeded, but not quenched, by the tip's magnetic field gradient.
The localized polarization profile in Fig.~\ref{Fig:DNP_spins_sketch2}(b) is consistent with the electron spin diffusion length being negligibly small but the nuclear spin diffusion length being close to the unperturbed bulk value.

{
\renewcommand{\arraystretch}{1.25}
\begin{table*}
\caption{Estimated sensitivity and imaging resolution achievable in an MRFM experiment with hyperpolarized magnetization.}
\begin{center}
\begin{tabular}{r|c|c|c|c|c|c|c|c|c|c}
Expt.
	& $N_{\text{s}}$
	& $\mu_{\text{s}}$
	& $G^{\text{tip}}_{zx}$
	& $p$
	& $0.49 p N_{\text{s}} F_1$
	& $P_{\delta F}$
	& $T_{\text{acq}}$
	& $F_{\text{min}}$
	& $\text{SNR}_2$
	& $\text{SNR}_2$
	\\
	
	&
	& $[\atto\newton \: \nano\meter/\milli\tesla]$
	& $[\milli\tesla/\nano\meter]$
	&
	& $[\atto\newton]$
	& $[\atto\newton^2/\hertz]$
	& $[\second]$
	& $[\atto\newton]$
	& (Eq.\ref{Eq:SNR2-symm:fast})
	& (Eq.\ref{Eq:SNR2-symm:slow})
	\\ \hline	
A
	& $1.9 \times 10^{5}$
	& $0.014$
	& $5.5$
	& $0.01$
	& $146$
	& $1600$
	& $60$
	& $5.2$
	& $1.1$
	& $28$
	\\

B
	& $4.0 \times 10^{3}$
	& $0.014$
	& $11.$
	& $0.18$
	& $110$
	& $100$
	& $60$
	& $1.3$
	& $3.5$
	& $85$
	\\

\end{tabular}
\end{center}
\label{Table:SNR}
\end{table*}
}

\paragraph*{Sensitivity and resolution}
The absolute proton spin polarization achieved here is far less than unity but nevertheless significant. At $B_0 = 0.568 \: \tesla$ and $T_0 = 4.2 \: \kelvin$, $p_{\text{therm}}= 1.58 \times 10^{-4}$.  Given the estimated enhancement of $\epsilon = 10$ to $20$, the absolute polarization after 15 seconds of DNP is $p = \epsilon \, p_{\text{therm}}= 1.6 \times 10^{-3}$ to $3.2 \times 10^{-3}$ (e.g., 0.16 to 0.32 percent). From simulations, we estimate the number of spins contributing to the signal to be $7.86 \; \times \; 10^{10}$ in a gradient of $G^{\text{tip}}_{zxx} \; = 1.1 \times 10^{-5} \; \milli\tesla \; \nano\meter^{-2}$. To determine the sensitivity of our experiment, we calculate an equivalent magnetic moment noise of \cite{Moore2009dec} $P_{\delta \mu} = 4 \; k_c^2 \; P_{\delta f} / f_c^2 (G^{\text{tip}}_{zxx})^2 = (2.6 \times 10^7 \; \mu_{\text{p}}/\sqrt{{\text{Hz}}})^2$ from the $\delta f_{\text{c}}^{\text{rms}} = 3.5 \; \milli\hertz$ frequency noise observed in a 0.25 Hz bandwidth in the Fig.~\ref{Fig:DNP_spins_sketch2}(c) experiment. As a point of comparison, given the estimated spin polarization of $3.2 \times 10^{-3}$ and proton density of 50 spins $\nano\meter^{-3}$, $2.6 \times 10^7 \; \mu_{\text{p}}$ is the net magnetic moment from $(550 \; \nano\meter)^3$ of spins.

The sensitivity and imaging resolution achievable in a small-magnet tip MRFM experiment with hyperpolarized magnetization  is estimated using Eqs.\ref{Eq:SNR2-symm:fast} and \ref{Eq:SNR2-symm:slow} in Table~\ref{Table:SNR}.
The table considers two example cases.
In Expt.~A, we assume the tip and surface noise from Ref.~\citenum{Longenecker2012nov} with the number of spins detected per point from Ref.~\citenum{Degen2009feb}.
We take the base electron polarization to be $0.1$, consistent with our present operating conditions of $18 \: \giga\hertz$ and $4.2 \: \kelvin$.
Assuming a DNP efficiency of 10 percent (e.g. $\epsilon = 66$) gives $p = 0.01$.
We calculate the SNR in two cases.
The Eq.~\ref{Eq:SNR2-symm:fast} calculation assumes $T_1 = 12 \: \second$ (\emph{vide supra}) and $\tau_{\text{m}} = 0.020 \: \second$ \cite{Degen2009feb} while the Eq.~\ref{Eq:SNR2-symm:slow} calculation assumes $\tau_{\text{m}} \rightarrow T_1$.
In Expt.~B we assume a smaller-radius tip having a gradient improved by a factor of two, a base electron polarization of $p = 0.60$ (e.g. $60 \: \giga\hertz$, $2.1 \: \kelvin$), a DNP efficiency of 30 percent (e.g. $\epsilon = 200$), an optimized force noise taken from Ref.~\citenum{Hickman2010nov}, and consider the same two $\tau_{\text{m}}$ cases.

In the virus imaging experiment of Degen and co-workers \cite{Degen2009feb} the peak lateral gradient was $G^{\text{tip}}_{zx} = 4 \: \milli\tesla \: \nano\meter^{-1}$; the peak spin variance signal, $\sigma^{2}_{\text{spin}} = 600 \: \atto\newton^2$, arose from approximately $N_{\text{s}} = \sigma^{2}_{\text{spin}} / (\mu_{\text{p}}^2 \: (G_{zx}^\text{tip})^2) = 1.9 \times 10^{5}$ protons in resonance.
The associated volume of spins in resonance is $V_{\text{s}} = 3800 \: \nano\meter^3$ (assuming a proton density of $50 \: \text{spins}/\nano\meter^3$).
The computed spin density at each sample point in the  Ref.~\citenum{Degen2009feb} experiment contains contributions from signal measured at many sample locations; due to this signal averaging, the imaging resolution is smaller than $V_{\text{s}}$ by nearly an order of magnitude, $4 \: \nano\meter \times 10 \: \nano\meter \times 10 \: \nano\meter = 400 \: \nano\meter^3$.
Let us take the $N_{\text{s}}$ used to compute the signal-to-noise ratios in Table~\ref{Table:SNR} and divide by ten to account for the improvement in the signal-to-noise ratio expected from the image-reconstruction step.
The implied imaging resolution ranges from slightly worse than $(6 \: \nano\meter)^3$ in the $\tau_{\text{m}} \ll T_1$ case of Expt.~A to much better than $(2 \: \nano\meter)^3$ in the $\tau_{\text{m}} \rightarrow T_1$ case of Expt.~B.
In both experiments, the SNR is improved 25 fold by pushing $\tau_{\text{m}}$ to the $T_1$ limit.

Creating, detecting, and imaging DNP-enhanced nuclear magnetization in a small-tip nanometer-resolution MRFM experiment will be challenging.
We can draw on the above analysis and prior work to enumerate the key expected challenges.

\paragraph*{Polarization}
The cross-effect DNP mechanism is a three-spin process requiring two electrons in close proximity having a difference in Larmor frequencies equal to the Larmor frequency $\omega$ of the adjacent nuclear spins.
Spin diffusion is needed to share the nuclear spin polarization created near the paramagnetic site with spins many nanometers away; to achieve homogeneous magnetization in a single macromolecule, it will therefore be important to keep the gradient below $G^{\text{crit}}_{2} =  0.7 \: \milli\tesla \: \nano\meter^{-1}$.
The gradient could impede CE-DNP in a more fundamental way as well, by creating a difference in Larmor frequency for adjacent electrons that could exceed $\omega = \gamma_{\text{p}} \: B_0$ and therefore shut down the cross-effect mechanism.
The associated critical gradient is
\begin{equation}
G^{\,\text{crit}}_{\text{CE-DNP}}
	= \frac{\gamma_n}{\gamma_e} \frac{B_0}{a}
	\label{Eq:GcritCEDNP}
\end{equation}
where $a$ is the distance between paramagnetic molecules.
For a paramagnetic dopant concentration of $40 \: \milli\text{M}$, the critical gradient is $G^{\,\text{crit}}_{\text{CE-DNP}} = 0.29 \: \milli\tesla \: \nano\meter^{-1}$ at $B_0 = 0.6 \: \tesla$ ($f_{\, \text{MW}} = 17 \: \giga\hertz$), rising to $G^{\,\text{crit}}_{\text{CE-DNP}} = 2.9 \: \milli\tesla \: \nano\meter^{-1}$ at $B_0 = 6 \: \tesla$ ($f_{\, \text{MW}} = 170 \: \giga\hertz$).

\paragraph*{Imaging}
Harnessing DNP-enhanced nuclear magnetization for nanometer-scale imaging, on the other hand, will require reducing or eliminating spin diffusion.
Assuming one-dimensional diffusion for simplicity, spin diffusion will broaden a $\sigma(0) = 1 \: \nano\meter$ wide Gaussian distribution of magnetization to a width of $\sigma(t) = 2 \: \nano\meter$ in only $t = 1.5 \sigma^2 / D = 3.3 \: \milli\second$.
This calculation indicates that nanometer-scale spin-magnetization gradients created during Fourier image encoding \cite{Kempf2003feb,Nichol2013sep} in an MRFM experiment will be erased rapidly by spin diffusion.
To avoid this problem, Kempf and Marohn have proposed applying rf pulses in synchrony with cantilever motion to enable FT-image encoding while eliminating spin evolution from dipolar couplings \cite{Kempf2003feb}.
Alternatively, an image can be acquired by collecting signal while slowly scanning the cantilever \cite{Degen2009feb}. To image a polarized-spin sample in this way, it will be important to keep magnetization from transferring between adjacent slices by, for example, detecting in a gradient larger than $G_2^{\text{crit}}$ to suppress spin diffusion.

\paragraph*{Detection}
For simplicity, here we used the dc CERMIT effect to observe the polarized nuclear spins. The signal-to-noise ratio in our experiments was limited by surface frequency fluctuations and not thermomechanical noise. It is likely that surface frequency noise will be an even more pronounced problem in a small-tip experiment \cite{Yazdanian2008jun,Yazdanian2009jun,Longenecker2012nov}. In future experiments, optimizing the signal-to-noise ratio will require balancing two competing demands. To achieve long $\tau_{\text{m}}$, the nuclear magnetization should be modulated slowly. To evade surface frequency noise, the nuclear magnetization should be modulated rapidly; at fast modulation frequencies parametric upconversion may be necessary to evade detector noise \cite{Moore2010jul}.
It is unknown how well this upconversion scheme will work in practice in a high surface noise environment.

\vspace{0.125in}

In an MRFM experiment, the per-spin sensitivity is proportional to the magnetic field gradient.  The gradient in the experiment of Ref.~\citenum{Longenecker2012nov} was $5.5 \: \milli\tesla \: \nano\meter^{-1}$, enabling $500 \: \mu_{\text{p}}$ sensitivity but already larger than both $G^{\text{crit}}_{2}$ and $G^{\,\text{crit}}_{\text{CE-DNP}}$; the gradient requirements for achieving optimized polarization, imaging, and high-sensitivity detection are seemingly incompatible.
Eberhardt, Meier, and coworkers encountered a similar problem when trying to perform NMR spectroscopy measurements in a sample-on-cantilever MRFM experiment \cite{Eberhardt2008oct}.
They obtained high-resolution spectra by cycling their millimeter-scale gradient source away from the sample temporarily to allow for a period of spin evolution under the chemical shift in a homogeneous field.
Shuttling the $200 \: \nano\meter$ wide tip of Ref.~\citenum{Longenecker2012nov} laterally by $500$ to $1000 \: \nano\meter$ on the timescale of $T_1$ (many seconds at cryogenic temperatures) appears feasible and would enable the independent optimization of the gradient during periods of polarization, image encoding, and spin detection.

\section{Conclusion}

Many magnetic resonance experiments have been shown to be compatible with magnetic resonance force microscopy.
These experiments include the Rabi nutation experiment \cite{Wago1996mar,Mamin2005jul}, spin echoes \cite{Wago1998jan}, dipolar spectroscopy \cite{Degen2005may,Degen2006sep}, cross polarization \cite{Lin2006apr}, indirect observation of low gamma nuclei via cross depolarization \cite{Eberhardt2007may}, two-dimensional spectroscopy \cite{Eberhardt2008oct}, and nuclear double resonance \cite{Poggio2009feb}.
Here we have used the widely applicable cross-effect DNP mechanism to create hyperthermal nuclear spin polarization in a thin-film polymer sample in a magnet-on-cantilever MRFM experiment.  If the challenges discussed above can be addressed, using DNP to create hyperthermal spin polarization in an MRFM experiment offers many exciting possibilities for increasing the technique's sensitivity in both imaging and double-resonance experiments.

\section*{Acknowledgements}

We acknowledge Doran D.\ Smith and the U.S. Army Research Laboratory (Adelphi, MD) for the use of a microwave source and amplifier; Ehsan Afshari (Cornell University) for assistance with Sonnet simulations; Doran D.\ Smith (U.S. Army Research Laboratory) and  H.\ John Mamin and Daniel Rugar (IBM Almaden Research Center) for sharing their cryogenic magnetic resonance force microscope designs with us; J.C.\ S\'{e}mus Davis (Cornell University) for aid in designing a vibration-isolation platform; Raffi Budakian and John M.\ Nichol (Univ.\ of Illinois, Urbana-Champaign) for sharing the design of their Pan-style walkers with us; and Lee E. Harrell (U.S.\ Military Academy at West Point) for suggesting a novel probe- and magnet-hoisting mechanism.  We acknowlege Rico A.R.\ Picone, Joe L. Garbini, and John A.\ Sidles (Univ.\ of Washington, Seattle) for stimulating discussions and for sharing results with us prior to their publication.

This research was funded by the National Institutes of Health (grant no.\ R01GM070012-07), the Army Research Office (grant nos.\ W911NF-12-1-0221 and W911NF-14-1-0674), and the National Science Foundation through Cornell's GK12 Program (grant no.\ NSF DGE-1045513; ``Grass Roots: Advancing education in renewable energy and cleaner fuels through collaborative graduate fellow/teacher/grade-school student interactions'').  This work made use of the Cornell Center for Materials Research Shared Facilities which are supported through the National Science Foundation MRSEC program (grant no.\ DMR-1120296). This work also was performed in part at the Cornell NanoScale Facility, a member of the National Nanotechnology Coordinated Infrastructure (NNCI), which is supported by the National Science Foundation (grant no.\ ECCS-115420819).

\section*{Appendix}

\renewcommand{\thefigure}{A\arabic{figure}}
\renewcommand{\thetable}{A\arabic{table}}
\renewcommand{\theequation}{A.\arabic{equation}}

\setcounter{figure}{0}
\setcounter{table}{0}
\setcounter{equation}{0}

Here we calculate the signal-to-noise ratio for detecting the average magnetization in a magnetic resonance force microscope (MRFM) experiment. As described in the text, we let each experiment begin with an average magnetization equal to zero.
After waiting a time $T$ for the sample to polarize, there is now a spin force acting on the cantilever whose magnitude is
\begin{equation}
	F_{\text{s}}^{\, \text{init}}
		= p \, N_{\text{s}} F_1 \, (1 - e^{-T/T_1}) .
\end{equation}
At a subsequent time $t$ the force acting on the cantilever is $F_{\text{s}}(t) =  F_{\text{s}}^{\, \text{init}} e^{-t/\tau_{\text{m}}}$.  The force signal is multiplied by a matched filter $e^{-t/\tau_{\mathrm{m}}}$, integrated, and the resulting integral is divided by the observation time $t$ to obtain the following estimate for the magnitude of the mean spin-force signal:
\begin{equation}
	F_{\text{s}}^{\, \text{avg}}
	= p \, N_{\text{s}} F_1 \, (1 - e^{-T/T_1})
		\frac{\tau_{\text{m}}}{2 t}
			(1 - e^{-2 t/\tau_{\text{m}}}).
\end{equation}

There are two contributions to the force noise acting on the cantilever: the environmental force noise, whose power spectrum is given by Eq.~\ref{Eq:PdF}, and the force noise arising from magnetization fluctuations, whose variance is given by Eq.~\ref{Eq:sigma2spin}.
The two noise sources are uncorrelated and we may therefore add their variances to obtain the total force-noise variance observed during one measurement period: $\sigma^2_{F,\,1} = P_{\delta F}/(4 t) + \sigma^{2}_{\text{spin}}$, where $b = 1/(4 t)$ is the noise-equivalent bandwidth of the matched exponential filter.
The measurement is repeated $N_{\text{avg}} = T_{\text{acq}}/(T+t)$ times and the resulting signals are averaged together.
As a result of the averaging, the noise variance will be reduced by a factor of $1/N_{\text{avg}}$ to $\sigma^2_{F,\,N_{\text{avg}}} = \sigma^2_{F,\,1}/N_{\text{avg}}$.
The resulting signal-to-noise expression SNR$_2$ = $F_{\text{s}}^{\text{avg}}/\sigma_{\text{F, }N_{\text{avg}}}^2$ is given by Eq.~\ref{Eq:SNR2-symm}.

As described in the text, we can calculate approximate expressions for the SNR in the detector-noise limit and in the spin-noise limit.
These two cases can be summarized as
\begin{align*}
\text{detector-noise limited:} & \: \: t \ll t_{\text{r}}^{\text{opt}} \\
\text{spin-noise limited:} & \: \: t \gg t_{\text{r}}^{\text{opt}}
\end{align*}

For $t \ll T$ (or, equivalently, when $\tau_{\text{m}} \ll T_1$), the detector-limited SNR is maximized by setting $T \approx 1.256 \, T_1$ and $t \approx 0.628 \, \tau_{\text{m}}$ giving
\begin{equation}
	\text{SNR}_{2\text{A}}
	\approx 0.48 \,
	p \, \sqrt{N_{\text{s}}} \,
		\sqrt{\frac{T_{\text{acq}}}{t_{\text{r}}^{\text{opt}}}}
		\sqrt{\frac{\tau_{\text{m}}}{T_1}}.
	\label{Eq:SNR2-detector-limit}
\end{equation}
 The spin-noise limited SNR is maximized by setting $T \approx 1.256 \, T_1$ and keeping $t \ll \tau_{\text{m}}$ to give an SNR expression
\begin{equation}
	\text{SNR}_{2\text{B}}
	\approx 0.64 \,
	p \, \sqrt{N_{\text{s}}} \,
		\sqrt{\frac{T_{\text{acq}}}{T_1}}.		
	\label{Eq:SNR2-spin-noise-limit}
\end{equation}

\section*{Supporting Information}

\renewcommand{\thefigure}{S\arabic{figure}}
\renewcommand{\thetable}{S\arabic{table}}
\renewcommand{\theequation}{S.\arabic{equation}}

\setcounter{figure}{0}
\setcounter{table}{0}
\setcounter{equation}{0}

\subsection{Frequency noise and equivalent force noise}

\begin{figure}[b]
\includegraphics[width=3.25in]{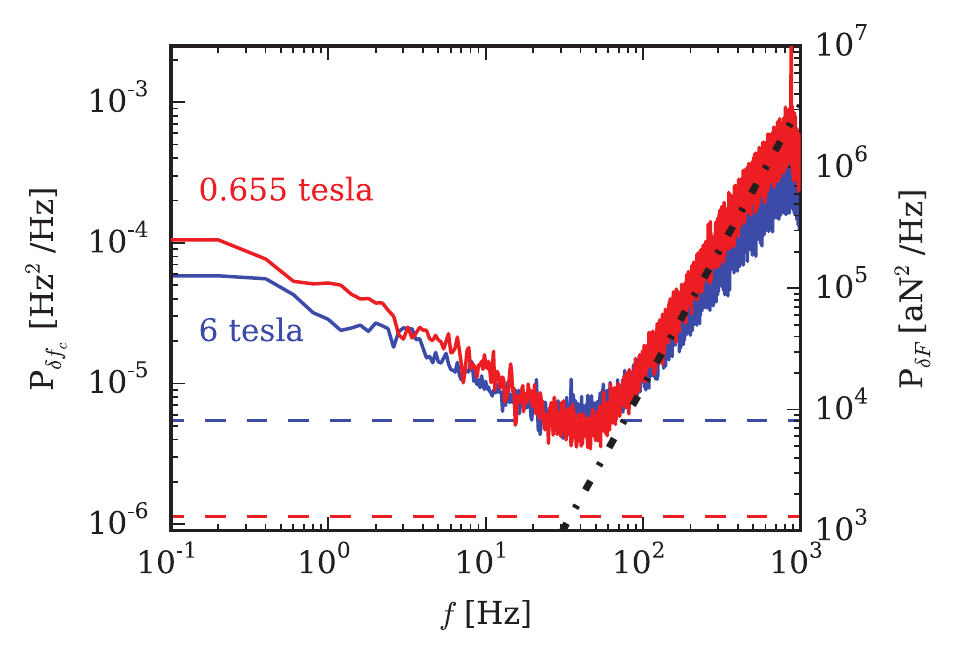}
\caption{Power spectrum of cantilever frequency fluctuations $P_{\delta f_{\text{c}}}$ versus offset frequency $f$ at $B_0 = 0.655 \: \tesla$ (red line) and $B_0 = 6.0 \: \tesla$ (blue line).  The $P_{\delta f_{\text{c}}}$ data above $f \geq 10^{2} \: \hertz$ was fit to Eq.~\ref{Eq:dfcdet} with $x_{\text{rms}} = 69 \: \nano\meter$ to obtain $P_{\delta x}^{\, \text{det}} = 3.6 \times 10^{-6} \: \nano\meter^2 \: \hertz^{-1}$ (dotted black line), the power spectrum of detector noise expressed in units of equivalent position noise.  The right-hand axis is $P_{\delta f_{\text{c}}}$ rewritten in terms of an equivalent force fluctuation using Eq.~\ref{Eq:Pdfavg-result} and $k_{\text{c}} = 1.0 \: \milli\newton \: \meter^{-1}$, $f_{\text{c}} = 3500 \: \hertz$, and $x_{\text{rms}}$.  The dashed lines are the thermo-mechanical force fluctuations calculated from Eq.~\ref{Eq:PdF-therm} at $B_0 = 0.655 \: \tesla$ (dashed red line), where the cantilever ringdown time $\tau_{\text{c}} = 0.94 \: \second$, and at $B_0 = 6.0 \, \tesla$ (dashed blue line), where  $\tau_{\text{c}} = 0.15 \, \second$.  Other experimental parameters: temperature $T_0 = 4.2 \: \kelvin$, tip-sample separation $h = 1500 \: \nano\meter$,  acquisition time $T_{\text{acq}} = 10  \, \second$ per average, and number of averages $n_{\text{avg}} = 32$. \label{Fig:noise}}
\end{figure}

In the experiments described in the manuscript, electron-spin resonance and nuclear magnetic resonance were registered as a change in the mechanical resonance frequency of a cantilever.
Thermo-mechanical position fluctuations place a fundamental limit on how small a cantilever frequency shift can be measured in a given averaging time \cite{Albrecht1991jan,Obukhov2007feb,Yazdanian2008jun,Dwyer2015jan}.
In this section we present cantilever frequency-fluctuation power spectra and use these spectra to assess how close the experiments in the manuscript were to operating at the thermo-mechanical limit.

A power spectrum of cantilever frequency fluctuations $P_{\delta f_{\text{c}}}(f)$ was collected at $B_0 = 0.655 \: \tesla$ and $B_0 = 6 \: \tesla$, in vacuum, at $4.2 \: \kelvin$ (see Fig.~\ref{Fig:noise}).  Apparent in the spectrum are $\propto 1/f$ dielectric fluctuations  at low offset frequency $f$ \cite{Yazdanian2009jun} and detector noise $\propto f^2$ at high $f$ \cite{Yazdanian2008jun}.  The detector-noise contribution to the cantilever frequency-noise power spectrum is \cite{Yazdanian2008jun,Dwyer2015jan}
\begin{equation}
P_{\delta f_{\text{c}}}^{\, \text{det}}(f)
 	= \frac{P_{\delta x}^{\, \text{det}}}{x_{\text{rms}}^2} \, f^2
	\label{Eq:dfcdet}
\end{equation}
where $x_{\text{rms}}$ is the root-mean-square cantilever amplitude and $P_{\delta x}^{\, \text{det}}$ is the power spectrum of detector noise expressed in units of equivalent position noise. The high-$f$ data in Fig.~\ref{Fig:noise} was fit to Eq.~\ref{Eq:dfcdet} to obtain $P_{\delta x}^{\, \text{det}} = 3.6 \times 10^{-6} \: \nano\meter^2 \: \hertz^{-1}$.

The cantilever frequency fluctuations can be analyzed, as follows, to obtain a power spectrum of equivalent force fluctuations.  Fluctuating forces acting on the cantilever lead to fluctuations in the cantilever position whose power spectrum is given by
\begin{equation}
P_{\delta x}(f)
	= \frac{P_{\delta F}(f)}{k_{\text{c}}^2}
	\frac{f_c^4}{(f_{\text{c}}^2 - f^2)^2 - f^2 f_{\text{c}}^2 / Q^2}
	\label{Eq:Px}
\end{equation}
where $P_{\delta F}(f)$ is the power spectrum of force fluctuations and $k_{\text{c}}$, $f_{\text{c}}$, and $Q$ are the cantilever spring constant, resonance frequency, and quality factor, respectively.  These fluctuations in cantilever position contribute noise to the measured cantilever frequency.  The resulting power spectrum of induced frequency fluctuations is given by
\begin{equation}
P_{\delta f_c}(f)
	= \frac{f^2}{2 \, x_{\text{rms}}^2}
	\left(
		P_{\delta x}(f + f_{\text{c}})
		+ P_{\delta x}(f - f_{\text{c}})
	\right) .
	\label{Eq:Pfc}
\end{equation}
We could at this point substitute Eq.~\ref{Eq:Px} into Eq.~\ref{Eq:Pfc} and obtain a relation between $P_{\delta f_c}$ to $P_{\delta F}$.  Before doing so, it is helpful to examine
\begin{multline}
P_{\delta x}(f \pm f_{\text{c}})
	= \frac{P_{\delta F}(f \pm f_{\text{c}})}{k_{\text{c}}^2} \\
	\times \frac{f_c^4}{(f_{\text{c}}^2 - (f \pm f_{\text{c}})^2)^2
		- (f \pm f_{\text{c}})^2 f_{\text{c}}^2 / Q^2} ,
		\label{Eq:Pdx-temp1}
\end{multline}
which simplifies to
\begin{equation}
P_{\delta x}(f \pm f_{\text{c}}) \approx
	\frac{P_{\delta F}(f \pm f_{\text{c}})}{k_{\text{c}}^2}
	\frac{f_{\text{c}}^4}{4 f^2 f_{\text{c}}^2 + f_{\text{c}}^4 / Q^2}
		\label{Eq:Pdx-temp2}
\end{equation}
where in going from Eq.~\ref{Eq:Pdx-temp1} to Eq.~\ref{Eq:Pdx-temp2} we have used that $f \ll f_{\text{c}}$.  Substituting Eq.~\ref{Eq:Pdx-temp2} into Eq.~\ref{Eq:Pfc} yields
\begin{multline}
P_{\delta f_c}(f)
	= \frac{f_{\text{c}}^2}{k_{\text{c}}^2 \, x_{\text{rms}}^2}
	\times
	\frac{1}{2}
	\left(
		P_{\delta F}(f_{\text{c}}+f) \right.
		\left. + P_{\delta F}(f_{\text{c}}-f)
	\right) \\
	\times
	\frac{f^2 f_{\text{c}}^2}{4 f^2 f_{\text{c}}^2 + f_{\text{c}}^4/Q^2}
	\label{Eq:Pdfc-temp1}
\end{multline}
where we have used that $P_{\delta F}(f)$ is an even function of $f$ to write $P_{\delta F}(f \pm f_{\text{c}}) \rightarrow P_{\delta F}(f_{\text{c}} \pm f)$.
This expression may be simplified further by defining
\begin{equation}
P_{\delta F}^{\, \text{avg}}(f_{\text{c}}, f)
	=
	\frac{1}{2}
	\left(
		P_{\delta F}(f_{\text{c}}+f)
		+ P_{\delta F}(f_{\text{c}}-f)
	\right),
	\label{Eq:PdFavg}
\end{equation}
the average power spectrum of force fluctuations at an offset frequency $f$ below and $f$ above the cantilever frequency.  Substituting Eq.~\ref{Eq:PdFavg} into Eq.~\ref{Eq:Pdfc-temp1} gives
\begin{equation}
P_{\delta f_c}(f)
	= \frac{f_{\text{c}}^2}{4 \, k_{\text{c}}^2 \, x_{\text{rms}}^2}
	P_{\delta F}^{\, \text{avg}} (f_{\text{c}}, f)
	\frac{1}{1 + f_{\text{c}}^2/(4 f^2 Q^2)} .
	\label{Eq:Pdfc-temp2}
\end{equation}
The last term in Eq.~\ref{Eq:Pdfc-temp2} becomes 1 in the limit that $f \gg f_{\text{c}}/(2 Q)$, that is, when $f$ is larger than the width of the oscillator resonance in $\text{cycles} \: \second^{-1}$.  In this limit,
\begin{equation}
P_{\delta f_c}(f)
	=  \frac{f_{\text{c}}^2}{4 \, k_{\text{c}}^2 \, x_{\text{rms}}^2}
	P_{\delta F}^{\, \text{avg}} (f_{\text{c}}, f) .
	\label{Eq:Pdfc-from-PdF}
\end{equation}
Solving for $P_{\delta F}^{\, \text{avg}}$ we obtain
\begin{equation}
P_{\delta F}^{\, \text{avg}} (f_{\text{c}}, f)
	= \frac{4 \, k_{\text{c}}^2 \, x_{\text{rms}}^2}{f_{\text{c}}^2}
	P_{\delta f_c}(f) .
	\label{Eq:Pdfavg-result}
\end{equation}
If the \emph{only} source of frequency noise was the underlying force noise, then we could use Eq.~\ref{Eq:Pdfavg-result} to calculate the fluctuating forces driving the cantilever from the measured power spectrum of cantilever frequency fluctuations.
In practice, however, $P_{\delta f_c}(f)$ contains additional contributions from surface noise and detector noise.
Applying Eq.~\ref{Eq:Pdfavg-result} to the measured frequency fluctuations we obtain an \emph{equivalent} or \emph{effective} power spectrum of force noise.
In this case Eq.~\ref{Eq:Pdfavg-result} can be interpreted as the power spectrum of force fluctuations  that, when applied to the cantilever, would yield frequency fluctuations having the observed power spectrum $P_{\delta f_{\text{c}}}$.
The $P_{\delta F}^{\, \text{avg}}$ calculated in this way is shown as the right-hand $y$ axis in Fig.~\ref{Fig:noise}.

For comparison, we can plot the power spectrum of thermo-mechanical force fluctuations $P_{\delta F}^{\, \text{therm}}$.  This power spectrum is independent of frequency.  In terms of measured parameters,
\begin{equation}
P_{\delta F}^{\, \text{therm}}
	= \frac{2 k_{\text{B}} T_0 k_{\text{c}}}
	       {\pi^2 f_{\text{c}}^2 \tau_{\text{c}}}
    \label{Eq:PdF-therm}
\end{equation}
where $k_{\text{B}}$ is Boltzmann's constant, $T_0$ is temperature, and $\tau_{\text{c}}$ is the cantilever ringdown time. Comparing the observed $P_{\delta F}^{\, \text{avg}}$ data in Fig.~\ref{Fig:noise} to the calculated $P_{\delta F}^{\, \text{therm}}$, we see that the equivalent force noise at $B_0 = 6.0 \: \tesla$ was near the thermo-mechanical limit at offset frequencies $20 \: \hertz < f < 50 \: \hertz$ while at $B_0 = 0.655 \: \tesla$ the equivalent force noise was never better than $10 \times$ the thermo-mechanical limit.

\subsection{Effect of cantilever motion on the DNP signal}

During the NMR, ESR, and DNP experiments described in the manuscript, the cantilever was oscillated at its resonance frequency.  According to Eq.~\ref{Eq:Pdfc-from-PdF}, cantilever root-mean-square frequency noise is inversely proportional to cantilever amplitude.  Moreover, in the ESR experiment some cantilever motion is required to scan the resonant slice through the sample and bring a measurably large number of electron spins into resonance.

\begin{figure}
\includegraphics[width=2.5in]{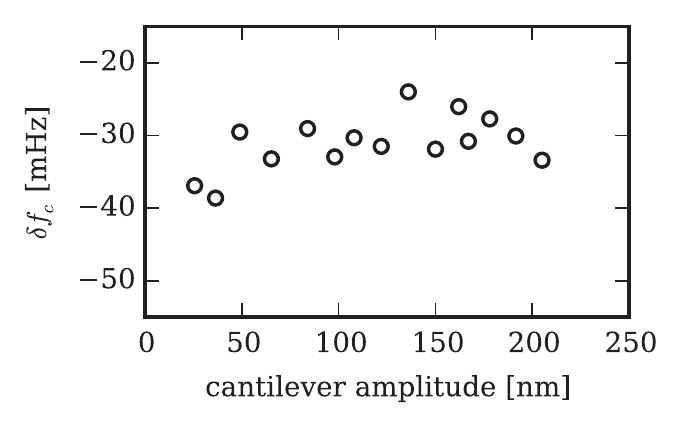}
\caption{DNP-enhanced $^{1}\text{H}$ spin signal versus peak-to-peak cantilever amplitude.  Experimental parameters: $B_0 = 0.655 \: \tesla$, $h = 1500 \: \nano\meter$, $f_{\text{MW}} = 18.5 \: \giga\hertz$, irradiation time $\tau = 20 \: \second$, $f_{\text{rf}} = 27.5 \: \mega\hertz$, $\Delta f_{\text{rf}} = 1 \: \mega\hertz$. }
\label{Fig:DNP_cant_amp}
\end{figure}

The microwave induced $^1$H spin signal was found to be independent of the peak-to-peak displacement of the cantilever as seen in Fig.~\ref{Fig:DNP_cant_amp}.  This observation is consistent with the manuscript's finding that the nuclear spins are polarized in a thin region on the proximal and distal sides of the resonant slice.  Oscillating the cantilever in the $x$ direction causes a lateral blurring of the resonant slice but does not, to first order, change location or thickness of the slice in the $z$ direction.  Consequently, increasing the oscillation amplitude of the cantilever should not cause any cancelation of the DNP enhancement.

\subsection{Coplanar waveguide details}

\begin{table}
\begin{ruledtabular}
\begin{tabular}{cccc}
frequency & parameter & measured & simulated \\ \hline
\\
$210 \: \mega\hertz$ & $S_{21}$ & $-1.1 \: \text{dB}$ & $-0.95 \: \text{dB}$ \\
$210 \: \mega\hertz$ & $S_{11}$ & $-19.0 \: \text{dB}$ & $-19.6 \: \text{dB}$ \\
\\
$17 \: \giga\hertz$ & $S_{21}$ & $-17 \: \text{dB}$ & $-1.2 \: \text{dB}$ \\
$17 \: \giga\hertz$ & $S_{11}$ & $-12 \: \text{dB}$ & $-13.2 \: \text{dB}$ \\
\end{tabular}
\end{ruledtabular}
\caption{Coplanar waveguide scattering parameters measured at room temperature in air and simulated using SONNET.\label{Table:SP}}
\end{table}

The coplanar waveguide consisted of two sections --- a copper CPW on an Arlon substrate and a copper CPW fabricated on high resistivity silicon.

{\bf Arlon section} --- The CPW-on-Arlon section was purchased from PCB Fab Express with precut holes to facilitate making a connection to an SMA coaxial connector.
The Arlon substrate had a thickness of $H = 2540 \: \micro\meter$ and a (specified) relative permittivity of $\epsilon_{\text{r}} = 9.8$.
The waveguide was made of $35 \: \micro\meter$ thick copper.
The waveguide's center line was $w = 457 \: \micro\meter$ wide and the gap to the flanking ground plane was $s = 228.6 \: \micro\meter$ wide on each side.
A $10 \: \milli\meter$ by $2 \: \milli\meter$ section was removed from the center of the Arlon substrate to accommodate the CPW-on-Si section described below.

{\bf Silicon section~} --- The CPW-on-Si section of the waveguide was microfabricated at the Cornell Nanoscale Science and Technology Facility.
The substrate was made of high-resistivity silicon, had a thickness of $H = 500 \: \micro\meter$, and had a (specified) relative permittivity of $\epsilon_{\text{r}} = 11.8$.
The waveguide was made of $0.2 \: \micro\meter$ thick copper.
The waveguide's outer center line was $w = 480 \: \micro\meter$ wide and the gap to the flanking ground plane was $s = 230 \: \micro\meter$ on each side.
This section tapered, over a distance of $450 \: \micro\meter$, to a narrower waveguide; the $w/s$ ratio was maintained in the tapered region.
The narrower, ``microwire'' section of coplanar waveguide was $L = 500 \: \micro\meter$ long, $w = 10 \: \micro\meter$ wide, and had an $s = 6 \, \micro\meter$ gap.
The dimensions of the $10 \: \milli\meter$ by $2 \: \milli\meter$ hole in the CPW-on-Arlon section were precisely measured, and the CPW-on-Si was cut using a dicing saw to fit into the hole leaving less than a $200 \: \mu\meter$ gap between the two sections.

{\bf Connections} --- The two CPWs were connected \emph{via} wire bonds. Three wire bonds were used to connect the center line and three wire bonds were used to connect each of the flanking ground planes.

{\bf $\bm{S}$ parameters} ---  See Table~\ref{Table:SP} for measured and calculated scattering parameters.

\subsection{Adiabaticity of nuclear spin inversions}

The CPW described above was designed to deliver broadband irradiation.
Electromagnetic simulations (Sonnet Software, Inc.) predicted a transverse magnetic field strength $B_1$ of $2.5 \: \milli\tesla$ with only $200 \: \milli\watt$ of input power at frequencies below $5 \: \giga\hertz$ where simulated and measured scattering parameters agreed within $1 \: \text{dB}$.
To invert the nuclear magnetization reversibly, the nuclear spin magnetization must stay aligned with the effective field in the rotating frame during an adiabatic rapid passage through resonance.
Maintaining this alignment requires a $B_1$ large enough to meet the adiabatic condition,
\begin{equation}
B_1^2 \gg \frac{1}{2\pi\gamma} \frac{d}{d t} \Delta B_0
\label{Eq:adiabatic_condition}
\end{equation}
with $d \Delta B_0/ d t$ the rate of change in the magnetic field and $\gamma = 42.56 \: \mega\hertz \: \tesla^{-1}$ the $^{1}\text{H}$ gyromagnetic ratio.
According to Eq.~\ref{Eq:adiabatic_condition}, a transverse magnetic field of strength $B_1 = 2.5 \: \milli\tesla$ should meet the adiabatic condition during a $\Delta f_{\text{rf}} = 1 \: \mega\hertz$ sweep as long as the sweep duration $\Delta t_{\text{rf}}$ was $\geq 0.014 \: \milli\second$.

Harrell {\it et al.} provide guidelines that allow us to further quantify how efficiently we are inverting nuclear spins~\cite{Harrell2004mar}.
Considering a spin-1/2 system and a finite radiofrequency sweep rate, one can calculate the probability that spins undergo a diabatic transition rather than an adiabatic transition using
\begin{equation}
P = \exp \left( \frac{- (2 \pi \gamma \ B_1)^2}{4 \, \lvert d \, f_{\text{rf}}/dt \rvert } \right) .
\label{eq:diabatic_prob}
\end{equation}
Under our experimental conditions, we calculate a 10$\%$ likelihood of a diabatic transition with a $\Delta f_{\text{rf}} = 1 \: \mega \hertz$ sweep lasting $\Delta t_{\text{rf}} = 0.021 \: \milli \second$.

Equation~\ref{eq:diabatic_prob} is valid in the limit that $\Delta f_{\text{rf}} \geq 5 \gamma B_1$.
This condition sets a lower limit on the width of an ARP frequency sweep necessary to prevent projection losses --- losses incurred from projecting the magnetization on the effective field when the rf is turned on.
For $B_1 = 2.5 \, \milli\tesla$, the 1 $\mega \hertz$ wide frequency sweep used throughout these experiments should be adequate.
The $\Delta f_{\text{rf}} = 0.3 \: \mega\hertz$ ARP sweeps used to map the DNP enhancement, however, do not strictly satisfy this condition.
The signal from the $0.3 \, \mega\hertz$ sweeps was likely affected by (modest) projection losses.

\begin{figure*}
\includegraphics[width=6.50in]{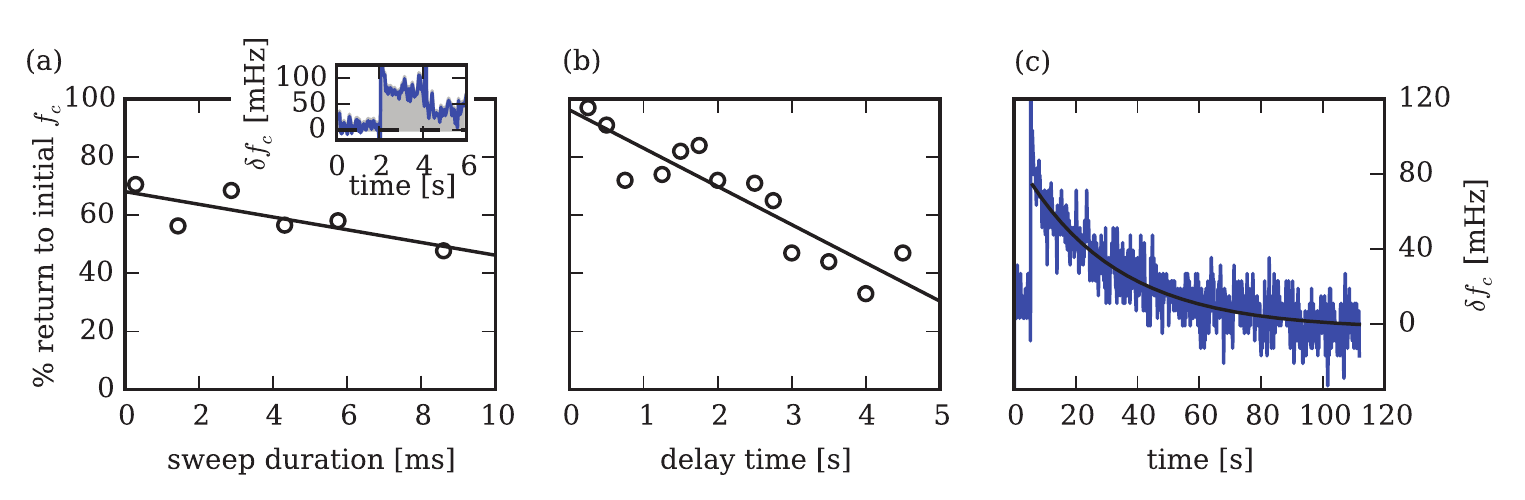}
\caption{The percent return of the cantilever resonance frequency following the application of two ARP sweeps applied to invert the sample's $^{1}\text{H}$ magnetization.  (a) Percent return versus the duration of the sweep, $\Delta t_{\text{rf}}$, with the delay between the sweeps fixed at $2 \: \second$.  (b) Percent return versus the delay between the sweeps, with the sweep duration fixed at $\Delta t_{\text{rf}} = 0.28 \: \milli\second$. (c) The cantilever frequency shift following a single $\Delta t_{\text{rf}} = 0.28 \: \milli\second$ duration sweep. The solid black line is a fit to an exponential decay with time constant $T_{1} = 30.9 \pm 0.9 \: \second$.  Experimental parameters: $B_0 = 4.93 \tesla$, $h = 1500 \: \nano\meter$, $f_{\text{rf}} = 210 \: \mega\hertz$, $\Delta f_{\text{rf}} = 1 \: \mega\hertz$, and $B_{1} = 2.5 \: \milli\tesla$ (estimated).  \label{Fig:spin_diffusion}}
\end{figure*}

The applied sweeps in our experiments were $0.28$ to $2.8 \: \milli \second$ in duration -- sufficient, we predicted, to meet the adiabatic condition with negligible diabatic transitions and projection losses.
After applying consecutive adiabatic rapid passage sweeps through resonance, however, we did not observe a complete return of the cantilever resonance frequency to its initial value (see the experiments and data presented in Fig.~\ref{Fig:spin_diffusion}).
Figure ~\ref{Fig:spin_diffusion}(a) shows the percent return of the cantilever frequency to its initial value following two identical ARP sweeps with a two second delay between them.
The fidelity of the inversions is poor.
Three hypotheses were developed to explain the observation:
\begin{enumerate}
\item  the oscillating field was not as strong as predicted; thus, we were not meeting the adiabatic condition;
\item a short $T_{1\rho}$ was causing a loss of magnetization during the rf sweep; and
\item spin diffusion was moving polarized spins out of the resonant slice during the delay before the second rf sweep was applied to re-invert the spins.
\end{enumerate}
As the duration of the sweep was shortened, the percent return improved, indicating that we were likely meeting the adiabatic condition but were possibly losing magnetization due to a short $T_{1\rho}$.  Fig.~\ref{Fig:spin_diffusion}(b) shows that the percent return improves as the inter-sweep delay is decreased.  Fig.~\ref{Fig:spin_diffusion}(c) shows a real-time measurement of the $^{1}\text{H}$ spin-lattice relaxation time; the measured $T = 30.9 \pm 0.9 \: \second$ is considerably longer than the few-second lifetime of the inverted magnetization apparent in Fig.~\ref{Fig:spin_diffusion}(b).  Taken together, these findings support the hypothesis that spin diffusion is carrying the inverted spin polarization away from the resonant slice on the time scale of just a few seconds.

\section{Absolute nuclear spin polarization}

\begin{figure*}
\begin{center}
\includegraphics[width=4.50in]{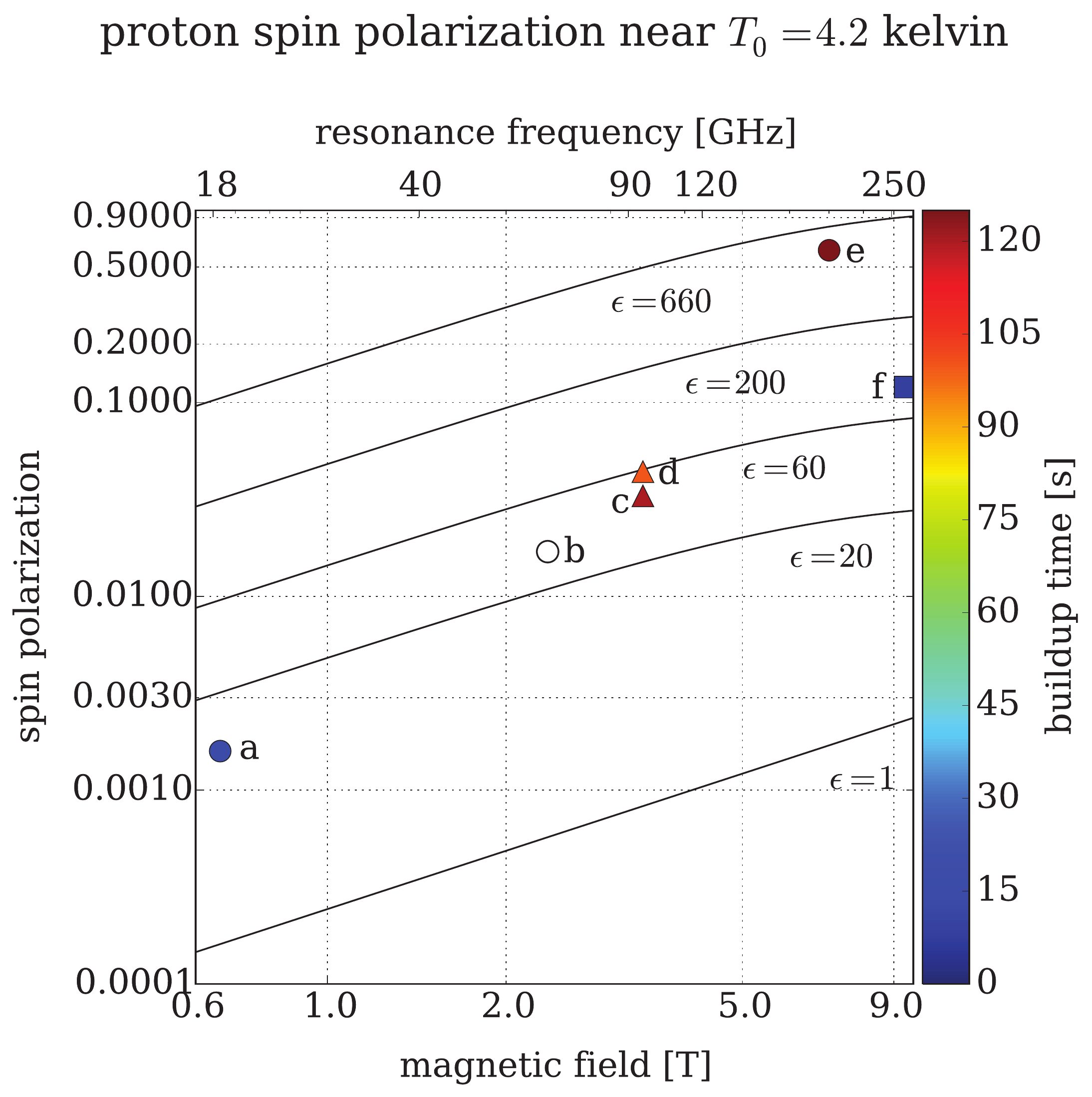}
\end{center}
\caption{Proton spin polarization achievable \emph{via} dynamic nuclear polarization at $T_0 = 4.2 \: \kelvin$ as a function of magnetic field, for representative enhancement factors ranging from $\epsilon = 1$ (no enhancement; lower curve) to $\epsilon = 660$ (full enhancement; upper curve). At full enhancement, the proton polarization is equal to the electron spin polarization.  Also plotted is the absolute $^{1}\text{H}$ polarization achieved in previous low-temperature DNP experiments: (a) this experiment, (b) Ref.~\citenum{Cho2007aug}, (c) Ref.~\citenum{Shimon2014apr}, (d) Ref.~\citenum{Shimon2012mar}, (e) Ref.~\citenum{Siaw2012aug}, and (f) Ref.~\citenum{Thurber2010jun2}.  The marker type indicates the operating temperature: circles for $T_0 = 4.2 \: \kelvin$, triangles for $T_0 = 6 \: \kelvin$, and squares for $T_0 = 7 \: \kelvin$.  The fill color indicates the nuclear magnetization buildup time; see the legend on the right-hand side of the plot.  The buildup time for experiment (b) is unknown. \label{Fig:proton_polarization_DNP_with_expt}}
\end{figure*}

The experiment described in this manuscript is the first time that microwave-assisted DNP has been definitively demonstrated in an MRFM experiment.
A 10 to 20-fold $^{1}\text{H}$ polarization enhancement was achieved.
Figure~\ref{Fig:proton_polarization_DNP_with_expt} summarizes the absolute polarization and buildup time achieved in inductively-detected DNP experiments carried out at temperatures ranging from $T_0 = 4.2 \: \kelvin$ to $T_0 = 7 \: \kelvin$.
The significantly greater enhancements achieved in these previous DNP experiments often came at the expense of long polarization buildup times.
The buildup time of $\tau \sim 13 \: \second$ seen in this experiment is favorably low compared to prior inductively-detected low temperature DNP experiments.
Implementing more optimized polarizing agents\cite{Cho2007aug,Shimon2014apr,Shimon2012mar,Siaw2012aug,Thurber2010jun2}, freezing the sample in a partially deuterated glass-forming solvent matrix, and operating at higher fields should lead to significantly greater absolute $^{1}\text{H}$ polarization in the MRFM experiment.

\subsection{Author contributions}


\noindent Short summary --- Author contributions: C.E.I.\ and J.A.M.\ designed the research;
E.W.M., J.L.Y., and L.C.\ designed and built the microscope;
L.C., J.L.Y., C.M.G., C.E.I. and H.L.N.\ cold-tested and characterized the microscope;
C.M.G. and P.T.N. fabricated coplanar waveguides;
P.T.N fabricated cantilevers;
C.E.I., H.L.N., and E.A.C.\ developed experimental protocols;
C.E.I.\ performed the research; and
C.E.I.\ and J.A.M.\ analyzed data and co-wrote the manuscript.


\noindent Long summary ---

\begin{itemize}

\item Corinne E.\ Isaac: cold-tested the microscope; characterized and optimized the Pan walkers; developed alignment protocols; designed and simulated the current version of the coplanar waveguide; performed experiments; analyzed data; and co-wrote manuscript

\item Christine M.\ Gleave: cold-tested the microscope; designed, simulated and fabricated initial versions of coplanar waveguides

\item Pam\'{e}la T.\ Nasr: fabricated the cantilever and CPW; fabricated grids for aligning the cantilever to the CPW

\item Hoang Long Nguyen: aided in cold-testing the microscope; helped develop alignment protocol

\item Elizabeth A.\ Curley: helped develop alignment protocol

\item Eric W.\ Moore: co-designed and co-built the microscope superstructure

\item Jonilyn L.\ Yoder: co-designed and co-built the microscope superstructure, cold-tested the microscope

\item Lei Chen: built the Pan-style walkers; designed the probe head; assembled and performed initial cold-testing of the probe head and microscope

\item John A.\ Marohn: designed experiments; analyzed data; and co-wrote manuscript

\end{itemize}


\providecommand*{\mcitethebibliography}{\thebibliography}
\csname @ifundefined\endcsname{endmcitethebibliography}
{\let\endmcitethebibliography\endthebibliography}{}

\label{TheEnd}

\end{document}